\documentstyle[12pt,amstex]{article}

\font\twelvemsb=msbm10 scaled 1200 
\def\Bbb#1{\hbox {\twelvemsb#1}}

\newcommand{\M}{{\cal M}}
\newcommand{\U}{{\cal U}}
\newcommand{\HH}{h}
\newcommand{\V}{{\cal V}}

\newcommand{\N}{{\cal N}}
\newcommand{\nn}{\Omega}

\newcommand{\Gzeta}{G^{\vec{\zeta}}}
\newcommand\bm[1]{\mbox{\boldmath$#1$}}
\newcommand\cm[1]{\bm{\cal#1}} 
\newcommand\sm[1]{{\mbox{\tiny \boldmath${#1}$}}}
\newcommand{\CH}{{\frak{X}}}
\newcommand{\Lam}{\Lambda^{1}}
\newcommand{\Txi}{T^{\vec{\xi}}}
\newcommand{\Tzeta}{T^{\vec{\zeta}}}
\newcommand{\T}{T^{\vec{\xi}}_{\sm{W}}}
\newcommand{\TzeW}{T^{\vec{\zeta}}_{\sm{W}}}
\newcommand{\g}{\T(g)}
\newcommand{\gzeta}{\TzeW(g)}
\newcommand{\eeta}{{\eta^{\prime}}}
\newcommand{\ric}{{R^{\prime}}}
\newcommand{\ji}{{\frak{S} (\U)}}
\newcommand{\C}{{ \cal C}} 
\newcommand{\R}{{\cal R}} 
\newcommand\ov[1]{\overline{#1}} 
\newcommand{\F}{{\cal F}}  
\newcommand{\FF}{\bm{\cal F}}

\newcommand{\I}{{\cm I}}
\newcommand{\Fsq}{\bm{{\cal F}^2}}
\newcommand{\Fprsq}{\bm{{{\cal F}^{\prime}}^2}}
\newcommand{\Fpr}{{{\cal F}^{\prime}}}  
\newcommand{\FFpr}{\bm{{\cal F}^{\prime}}}  
\newcommand{\ssig}{{\sigma^{\prime}}}
\newcommand{\sssig}{{\sigma^{\prime \prime}}}
\newcommand{\Y}{{\cal Y}}
\newcommand{\Ypr}{{\Y^{\prime}}}

\newcommand{\Apr}{{A^{\prime}}}
\newcommand{\cpr}{{c^{\prime}}}
\newcommand{\nablapr}{{\nabla^{\prime}}}
\newtheorem{lemma}{Lemma}
\newtheorem{theorem}{Theorem}

\newtheorem{proposition}{Proposition}
\begin{document}

\title{Spacetime Ehlers group: Transformation law for the Weyl
tensor}
\author{Marc Mars \\Albert Einstein Institut \\
Max Planck Institut f\"ur Gravitationsphysik, \\
Am M\"uhlenberg 1, D-14476 Golm, Germany.}
\date{5 January 2001}
\maketitle
\begin{abstract}

The spacetime Ehlers group, which is a symmetry of
the Einstein vacuum field equations for strictly stationary spacetimes,
is defined and analyzed in a purely
spacetime context (without invoking the projection formalism).
In this setting, the Ehlers group finds its natural description
within an infinite dimensional group
of transformations that maps Lorentz metrics into Lorentz metrics and
which may be of independent interest. The Ehlers group is shown
to be well defined independently of the causal character of the Killing
vector (which may become null on arbitrary regions). We analyze which 
global conditions are required on the spacetime for
the existence of the Ehlers group. The transformation law for 
the Weyl tensor under Ehlers transformations is explicitly obtained.
This allows us to study  where, and under
which circumstances, curvature
singularities in the transformed spacetime will arise.
The results of the paper are applied to obtain a local
characterization of the Kerr-NUT metric. 

\end{abstract}


\newpage

\section{Introduction.}

Strictly stationary spacetimes (i.e. spacetimes
admitting a Killing vector which is timelike everywhere)
can be conveniently studied by using the projection formalism
introduced by Geroch \cite{Geroch}, which consists in factoring out
the action of the Killing field by projecting all spacetime objects
onto the set of trajectories of the Killing vector. This method has been
extensively used specially because Einstein's field
equations simplify notably in this formalism
(see \cite{BeigSchmidt}
for a recent review on time independent gravitational fields). It
 was used, for instance, to find that 
Einstein's vacuum field equations for stationary spacetimes admit 
a finite dimensional symmetry group, (i.e. a group of transformations that
maps solutions into solutions). This is the so-called Ehlers group
\cite{Ehlers},\cite{Geroch}, which has been applied to many problems, 
ranging from the discovery of new solutions (see e.g. \cite{KSMH})
to the proof that the vacuum field equations in the stationary and axially
symmetric case form an integrable system (\cite{Geroch2} and e.g.
\cite{Hoenselaers} and references therein). 

Despite the power of the projection formalism, there are
circumstances in which it cannot be applied. For instance,
the set of trajectories of the Killing vector may fail to be a smooth
manifold. This is no problem when only local considerations are relevant
because for any point in the spacetime, there always exists a 
sufficiently small neighbourhood of it such that the quotient set
is a smooth manifold. However, for problems involving global aspects 
the question does become important and one should
investigate whether the set of trajectories is a smooth manifold. This
is, in general, difficult 
(see however \cite{Harris} for a set of necessary conditions for the
quotient to be a smooth manifold). A second, and perhaps more important, 
shortness of the projection formalism is that stationary spacetimes may
develop ergospheres and horizons. At points where
the Killing vector becomes null, the metric in the manifold of
trajectories becomes degenerate and the projection formalism cannot be used.
This makes this formalism unsuitable for studying most of the
problems concerning stationary rotating black holes (in particular,
their uniqueness properties) because the 
whole domain of outer communication cannot be covered by the method.
A similar problem arises for
some rapidly rotating objects, which may also have ergorpheres in their
exterior (the rotating disk of dust \cite{Meinel}, for instance, shows
this behaviour). 

Thus, some other method must be used to analyze problems in which the 
projection formalism cannot be used.
A  natural approach is to work directly on the spacetime
and use only spacetime objects. In general, this is more
difficult because the action of the
Killing vector has not been factored out and the very existence
of the Killing vector still needs to be imposed on the spacetime.
Nevertheless, this method can
be used in all situations where the projection formalism fails.
Furthermore, working directly on the
spacetime gives sometimes new insights into the problem. This method have been
used recently to obtain local \cite{Mars1} and semilocal  \cite{Mars2}
characterizations of the Kerr metric 
which hold everywhere (including the ergorsphere and/or the
black hole region) and which involve spacetime objects only.

The Ehlers group mentioned above was defined within the projection
formalism and is known to map locally a strictly
stationary vacuum solution into another strictly stationary vacuum solution
(locally means that there exists a suitably small open neighborhood
of any point 
where the Ehlers transformation can be defined). This is generally
sufficient
for generating new vacuum solutions, 
because one can apply the transformation locally and, if desired and possible,
the transformed spacetime can be extended to a maximal solution.
However, there are other problems in which understanding the global
properties of the action of the Ehlers group is important. For instance,
there are approaches 
\cite{Gibbons}, \cite{Maison} for proving uniqueness theorems
for stationary black holes which make use of the Ehlers transformation (and its
generalization to other  non-linear sigma models). However, this can only be
justified as long as the global properties
of the Ehlers transformation can be
controlled. Thus, studying in detail the global requirements 
for the existence of the Ehlers transformation becomes necessary.
As discussed above, the projection formalism is not suitable to analyze
this kind of problems. Furthermore, the Ehlers group is defined only on
regions where the Killing vector is timelike or spacelike
and it is not clear  
a priori whether it can be smoothly extended through 
ergospheres or horizons. Explicit examples
suggest that this extension can be performed, but no general proof
has been given. To answer these questions we need to 
define and analyze the Ehlers group in a framework that avoids using the
projection formalism.

In this paper we perform a
detailed study of the Ehlers transformation in a spacetime setting. 
This will allow us to prove, first of all, that the Ehlers transformation is
well-defined at points where the Killing vector is null,
as one could have expected.
More interestingly, the spacetime approach will reveal 
several properties of the Ehlers group 
which are hidden in the quotient description. In particular,
we will show that the Ehlers group finds its natural description
within an infinite dimensional group of transformations which maps Lorentzian
metrics into Lorentzian metrics. The general form of the transformation is
\begin{eqnarray}
g^{\prime}_{\alpha\beta} \equiv  \nn^2 g_{\alpha\beta}
 - \zeta_{\alpha} W_{\beta} - \zeta_{\beta} W_{\alpha} - \lambda \nn^{-2}
W_{\alpha} W_{\beta},
\label{forma}
\end{eqnarray}
where $\zeta^{\alpha}$ is an arbitrary vector field,
$\lambda = - \zeta^{\alpha} \zeta_{\alpha}$ and $W_{\alpha}$ is an arbitrary
one-form constrained to satisfy $\Omega^2 = 1 + \zeta^{\alpha} W_{\alpha} > 0$.
The transformation (\ref{forma}) includes as particular cases the
Kerr-Schild transformation \cite{KS} 
(by setting $W_{\alpha} \propto \zeta_{\alpha}$
and $\zeta^{\alpha} \zeta_{\alpha} =0$ everywhere) and a
transformation put forward and studied by Bonanos \cite{Bona} 
(when $\zeta^{\alpha} W_{\alpha}=0$ and hence $\Omega^2=1$).
Since  (\ref{forma}) contains the Kerr-Schild transformation as a particular
case, it also allows for a generalization of
Kerr-Schild symmetries, which have been recently defined and
studied in \cite{CSS}.
This issue, however, will not be considered further in this paper.

The set of transformations (\ref{forma})
will be shown to form a group and 
its basic properties will be discussed. Obviously,
this full group does not map, in general, vacuum solutions into vacuum
solutions but it is likely that
suitable subsets of it (besides the Ehlers group) do have this property.
Using the spacetime description will allow us to discuss the necessary
and sufficient conditions for the Ehlers transformation
to be globally defined. Related to this question, 
the existence and location of curvature singularities in the transformed
spacetime will be studied. This will be done by obtaining 
an {\it explicit} expression for the
Weyl tensor of the transformed spacetime in terms of the original one.
The transformation law turns out to be is surprisingly simple and clear. 
Thus, 
the full geometry of the transformed metric can be determined without
having to
perform the Ehlers transformation explicitly. This may be
particularly interesting for stationary and axially symmetric spacetimes
where the Ehlers group
extends to an infinite dimensional group, the so-called Geroch group, which
can be understood as an iteration of Ehlers transformations with respect
to different Killing vectors. The results of this paper can also be
applied to that situation. 

The transformation law for the Weyl tensor will be applied
to find a local
characterization of the Kerr-NUT spacetime. First, we shall obtain
the simplest subset of vacuum solution which is invariant
under Ehlers transformations. Its defining property
turns out to be closely connected to the characterization 
of the Kerr metric found in \cite{Mars1}. This indicates once again 
that the Kerr metric enjoys a very privileged geometric position
because it can be characterized by a property which 
arises also naturally from the Ehlers group
(and therefore directly from the
underlying structure of Einstein vacuum field equations with a Killing
vector).

The paper is organized as follows. In section 2 we write down several
identities which are valid for any spacetime admitting a Killing vector,
irrespectively of its causal character. They
are useful for any four-dimensional description of spacetimes 
with a Killing vector (no field equations are assumed in this section).
While some of these equations are well-known, others appear to be new.
In section 3 we introduce an infinite dimensional
group of transformations which maps Lorentzian metrics into 
Lorentzian metrics and we discuss its basic properties.
In section 4 we introduce the Ehlers group as a particular case of this
infinite dimensional group of transformations. Then, we discuss which are the
requirements for the Ehlers transformation to exist globally and we prove
that vacuum solutions are mapped into vacuum solutions irrespectively of the causal
character of the Killing vector. This shows that the Ehlers transformation is a
symmetry of the vacuum field equations 
independently of whether the Killing vector has ergoregions and/or horizons.
In section 5 we make use of the identities in section 2 in order to obtain 
the transformation law for the Weyl tensor
under Ehlers transformations. The result is surprisingly simple and
elegant. Having obtained the form of
the transformed Weyl tensor,  we can identify  where and under
which circumstances curvature
singularities in the transformed spacetime will occur.
Finally, in section 6 we identify 
the simplest subset of stationary vacuum solutions 
which is invariant under Ehlers transformations.
We classify the orbits of the Ehlers group in this invariant subset.
The paper concludes
with a spacetime characterization of the Kerr-NUT metric, which is a direct
consequence of the results in this paper combined with the results in \cite{Mars1}.

\section{General Identities for Spacetimes with a Killing vector.}

In this paper, a $C^n$ spacetime denotes a paracompact, Hausdorff,
connected $C^{n+1}$ four-dimensional manifold endowed with a $C^n$ metric of
signature $(-1,1,1,1)$. All spacetimes are assumed to
be oriented with metric volume form $\eta_{\alpha\beta\gamma\delta}$.
$(\M,g)$ will denote
a $C^2$ spacetime admitting a $C^2$ Killing vector field $\vec{\xi}$.
The norm and twist one-form
of $\vec{\xi}$ are defined respectively
by $\lambda = - \xi^{\alpha} \xi_{\alpha}$ and 
$\omega_{\alpha} = \eta_{\alpha\beta\gamma\delta} \xi^{\beta} \nabla^{\gamma}
\xi^{\delta}$. In order to study 
spacetimes with a Killing vector of arbitrary causal character, it is
useful to employ self-dual 2-forms, which are complex
2-forms ${\cm{B}}$ satisfying ${\cm{B}}^{\star} = -i {\cm{B}}$,
where $\star$ is the Hodge dual operator. 
Our notation for $p$-forms is as follows.
Boldface characters are used for $p$-forms, non-boldface characters
are used for its components. For 
self-dual 2-forms, curly characters will be used (boldface for the $2$-form and
non-boldface for its components).

The 2-form 
$F_{\alpha\beta} \equiv \nabla_{\alpha} \xi_{\beta}$ and
its self-dual associate   $\F_{\alpha\beta} \equiv
F_{\alpha\beta} + i F^{\star}_{\alpha\beta}$
will play a fundamental role in the following. The 2-form
$\FF \equiv \frac{1}{2} \F_{\alpha\beta} dx^{\alpha} \wedge dx^{\beta}$ will
be called {\it Killing form} throughout this paper. 
The Ernst one-form $\bm{\sigma} = \sigma_{\mu} dx^{\mu}$
associated to $\vec{\xi}$ is defined by 
\begin{eqnarray}
\sigma_{\mu} \equiv 2 \xi^{\alpha} \F_{\alpha
\mu} = \nabla_{\mu} \lambda - i \omega_{\mu}.
\label{defsigma}
\end{eqnarray}
Two well-known (see e.g. \cite{Israel}) properties which are
valid for any self-dual 2-forms
$\bm{\cal X}$ and $\bm{\cal Y}$ are
\begin{eqnarray}
 {\cal X}_{\mu\sigma}  {\cal Y}_{\nu}^{\,\,\,\sigma} +  {\cal
Y}_{\mu\sigma}  {\cal X}_{\nu}^{\,\,\,\sigma}  = \frac{1}{2} g_{\mu\nu}
 {\cal X}_{\alpha\beta}  {\cal Y}^{\alpha\beta},
\hspace{1cm} 
 {\cal X}_{\alpha\beta}  {Y}^{\alpha\beta} = \frac{1}{2}
 {\cal X}_{\alpha\beta}  {\cal Y}^{\alpha\beta}.
\label{chichi}
\end{eqnarray}
where  $\bm{Y} = \mbox{Re} ( \bm{\cal Y})$ is the real part of
$\bm{\cal Y}$.
We now obtain some algebraic identities for $\FF$.
Directly from the definition of $\bm{\sigma}$
and the first equation in (\ref{chichi}) we get
\begin{eqnarray}
\sigma_{\alpha} \sigma^{\alpha} = -\lambda \Fsq,
\end{eqnarray}
where $\Fsq \equiv \F_{\alpha\beta} \F^{\alpha\beta}$. 
An important identity is
 \begin{eqnarray}
\label{xiF}
\eta_{\alpha\beta\mu\nu} \xi^{\mu} \F^{\rho\nu} =
- i \xi^{\rho}  \F_{\alpha\beta} + \frac{i}{2} \delta^{\rho}_{\alpha}
\sigma_{\beta} - \frac{i}{2} \delta^{\rho}_{\beta} \sigma_{\alpha},
\end{eqnarray}
which can be proven  by inserting 
$\F^{\rho\nu} = \frac{1}{2} i \eta^{\rho\nu}_{\,\,\,\,\,\,\,\gamma\delta}
\F^{\gamma\delta}$ into the left-hand side and expanding the products
of $\eta$'s. This identity also holds for an arbitrary self-dual two-form 
$\cm{X}$ provided $\bm{\sigma}$ is defined accordingly (see (\ref{defsigma})).
Another identity which will be useful in Sect. 5 is
\begin{eqnarray}
\sigma_{\beta} \nabla_{\alpha} \lambda + \sigma_{\alpha} \nabla_{\beta}
\lambda - g_{\alpha\beta} \sigma^{\mu} \nabla_{\mu} \lambda
+2 \xi_{\beta} \sigma^{\mu} F_{\alpha\mu} 
+ 2 \xi_{\alpha} \sigma^{\mu} F_{\beta\mu} - 4 \lambda
F_{\alpha}^{\,\,\,\mu} \F_{\mu\beta} = \hspace{1cm} \label{qq1} \\
\hspace{4cm} = \sigma_{\alpha} \sigma_{\beta}
+ \Fsq \left( \lambda g_{\alpha\beta} + \xi_{\alpha} \xi_{\beta} \right ).
\nonumber 
\end{eqnarray}
This expression can be proven by splitting the real and imaginary parts.
The real part of (\ref{qq1})  was already proven
in \cite{Mars1}. The imaginary part is easily shown by
using the first  identity in (\ref{chichi}) 
(with $\bm{\cal X} = \bm{\cal Y} = \FF$).  

Let us consider next identities involving covariant derivatives
of the Killing form and/or of the Ernst one-form. They already involve the
curvature of the spacetime. From 
$\nabla_{\mu} \nabla_{\alpha} \xi_{\beta} = \xi^{\nu}
R_{\nu \mu \alpha \beta}$, which is a well-known consequence of the
Killing equations, the following identity follows
\begin{eqnarray}
\nabla_{\mu} \F_{\alpha \beta} = \xi^{\nu} \R_{\nu\mu\alpha\beta},
\label{funda}
\end{eqnarray}
where $\R_{\nu\mu\alpha\beta}$ is the so-called {\it right self-dual
Riemann tensor} defined by
$\R_{\nu\mu\alpha\beta} = R_{\nu\mu\alpha\beta} + \frac{i}{2} 
\eta_{\alpha\beta\rho\sigma} R_{\nu\mu}^{\,\,\,\,\,\,\,\, \rho\sigma}$.
Some well-known properties of  $\R_{\alpha\beta\gamma\delta}$ are
\begin{eqnarray}
g^{\alpha\lambda} \R_{\alpha\beta\lambda\mu}  = R_{\beta\mu},
\hspace{1cm} \R_{\alpha\beta\lambda\mu} +
\R_{\alpha\lambda\mu\beta} +
\R_{\alpha\mu\beta\lambda} = i \eta_{\gamma\beta\lambda\mu} R^{\gamma}_{\,
\,\alpha}, \label{Bianchi1}\\
\R_{\alpha\beta\lambda\mu} - \R_{\lambda\mu\alpha\beta}
= i \left ( \eta_{\lambda\mu\alpha\sigma} R^{\sigma}_{\,\, \beta}
- \eta_{\lambda\mu\beta\sigma} R^{\sigma}_{\,\,\alpha} -
\frac{R}{2} \eta_{\lambda\mu\alpha\beta} \right ), \\
\nabla^{\alpha} \R_{\nu\alpha\beta\mu} =
\nabla_{\mu} R_{\nu\beta} - \nabla_{\beta} R_{\nu\mu} + i \eta_{\beta
\mu\rho\sigma} \nabla^{\sigma} R^{\rho}_{\,\,\nu},
\end{eqnarray}
where $R_{\alpha\beta} \equiv R^{\mu}_{\,\,\,\,\alpha\mu\beta}$
is the Ricci tensor and $R$ is the scalar of curvature.
From the definition of the Ernst one-form
and (\ref{funda})  we easily find 
\begin{eqnarray}
\nabla_{\alpha} \sigma_{\beta} - 2 \nabla_{\alpha} \xi^{\mu} \F_{\mu\beta}
= 2 \xi^{\mu} \xi^{\nu} \R_{\nu\alpha\mu\beta},
\label{derivsigma}
\end{eqnarray}
which will be of fundamental importance in Section 5. 
We now obtain identities for the divergence and the exterior derivative of the
Ernst one-form. Using
(\ref{derivsigma}) and of the properties of
$\R_{\alpha\beta\gamma\delta}$, we get
\begin{eqnarray}
\nabla_{\alpha} \sigma^{\alpha} = - \Fsq + 2 \xi^{\alpha} \xi^{\beta}
R_{\alpha\beta},
\label{divsigma} \\
\nabla_{\alpha} \sigma_{\beta} - \nabla_{\beta} \sigma_{\alpha}
= 2 i \, \xi^{\nu} \eta_{\nu\beta\mu\alpha} R^{\mu}_{\,\,\rho}
\xi^{\rho}. \label{dsigma}
\end{eqnarray}
Similarly, identities for the exterior derivative and the divergence
of the Killing form $\FF$ can be obtained directly from
(\ref{Bianchi1}) and the fundamental equation (\ref{funda}).
The results are
\begin{eqnarray}
\nabla_{\mu} \F_{\alpha\beta} + \nabla_{\alpha} \F_{\beta\mu}
+ \nabla_{\beta} \F_{\mu\alpha} = i \eta_{\gamma\mu\alpha\beta} 
\xi^{\nu} R^{\gamma}_{\,\,\nu}, \hspace{1cm}
\nabla_{\mu} \F^{\mu}_{\,\,\,\,\beta} = - \xi^{\nu} R_{\nu\beta}.
\label{cod}
\end{eqnarray}
These expressions prove the following well-known Lemma, which will be needed
below.
\begin{lemma} 
\label{lemmaone}
Let $(\M,g)$ be a smooth spacetime admitting a Killing vector 
$\vec{\xi}$ and let $\FF$ be its associated Killing form. Then,
the necessary and sufficient condition for $\FF$ to be closed 
is that $R_{\alpha\beta} \xi^{\beta} = 0$.
\end{lemma}

Finally, we write down identities for the covariant Laplacian of
$\F_{\mu\nu}$ and $\sigma_{\mu}$, i.e.
$\nabla_{\alpha} \nabla^{\alpha} \F_{\mu\nu}$
and $\nabla_{\alpha} \nabla^{\alpha} \sigma_{\mu}$. 
Since expanding the second covariant
derivatives in these expressions would lead to 
a rather long calculation, we recall
the well-known Weitzenb\"ock formula (see e.g. \cite{CBD})
which  relates the covariant Laplacian and the Hodge-Laplace operator 
$\bm{\Delta} \equiv \bm{d} \bm{\delta} + \bm{\delta} \bm{d}$
acting on $p$-forms, where
$\bm{d}$ is the exterior differential and $\bm{\delta}$ the codifferential
$\bm{\delta} = (-1)^p \star^{-1} \bm{d} \, \star$. For any $p$-form
$\bm{\Omega}$ we have
\begin{eqnarray}
(\bm{\Delta} \bm{\Omega} )_{\alpha_1 \cdots \alpha_p}
= - \nabla^{\mu} \nabla_{\mu}  {\Omega}_{\alpha_1 \cdots \alpha_p}
- \sum_{q=1}^{p} \left (-1 \right )^q R^{\beta}_{\,\,\alpha_q}
 {\Omega}_{\beta\alpha_1 \cdots \hat{\alpha}_q \cdots \alpha_p}
+ \hspace{25mm} \nonumber \\
\hspace{25mm} + 2 \sum_{r < q} \left (-1 \right )^{r+q} R^{\beta\,
\,\,\,\gamma}_{\,\,\,\alpha_r \,\,\,\alpha_q} \Omega_{\beta\gamma\alpha_1
\cdots \hat{\alpha}_{r} \cdots \hat{\alpha}_{q} \cdots \alpha_p}.
\label{Weiz}
\end{eqnarray}
$\bm{\Delta \FF}$ is easily calculated from
(\ref{cod}) after using the general equation $(\bm{\delta}
\bm{\Omega})_{\alpha_2
\cdots \alpha_p} =  - \nabla^{\mu}   {\Omega}_{\mu \alpha_2 \cdots \alpha_p}$.
The result is
\begin{eqnarray*}
(\bm{\Delta}
\FF)_{\alpha\beta} = \nabla_{\alpha} \left (\xi^{\nu} R_{\nu \beta} 
\right ) - \nabla_{\beta} \left ( \xi^{\nu} R_{\nu\alpha} \right )
- i \eta_{\alpha\beta\rho\sigma} \nabla^{\sigma} \left (
\xi^{\nu} R^{\rho}_{\,\,\nu} \right ).
\end{eqnarray*}
With this expression at hand, the covariant Laplacian of $\F_{\alpha\beta}$
follows from (\ref{Weiz}). The result takes a simpler
form if the Riemann tensor is decomposed in terms of the Ricci tensor 
and the Weyl tensor $C_{\alpha\beta\gamma\delta}$. In particular, after
defining the {\it self-dual
Weyl tensor} as 
$\C_{\nu\mu\alpha\beta} = C_{\nu\mu\alpha\beta} + \frac{i}{2} 
\eta_{\alpha\beta\rho\sigma} C_{\nu\mu}^{\,\,\,\,\,\,\,\, \rho\sigma}$, the
identity becomes
\begin{eqnarray}
\nabla_{\alpha} \nabla^{\alpha} \F_{\mu\nu} =
- \frac{1}{2} \F^{\alpha\beta} \C_{\alpha\beta\mu\nu}
+ \nabla_{\nu} \left (\xi^{\delta} R_{\delta\mu} \right )
- \nabla_{\mu} \left (\xi^{\delta} R_{\delta\nu} \right )
+ i \eta_{\mu\nu\rho\sigma} \nabla^{\sigma} \left (\xi^{\delta}
R_{\delta}^{\,\,\,\rho} \right ) + \frac{R}{3} \F_{\mu\nu}.
\label{covLapF}
\end{eqnarray}
Finally, we  evaluate
$\nabla_{\alpha} \nabla^{\alpha} \sigma_{\mu}$. In this case 
the calculation cannot be simplified  by evaluating  
$\bm{\Delta \sigma}$ first  because we do not have an identity 
for $\nabla_{\alpha} \Fsq$ yet (see (\ref{divsigma})).
So, we use the definition
$\sigma_{\alpha}= 2 \xi^{\beta} \F_{\beta\alpha}$ and expand the
derivatives explicitly. A somewhat long calculation
using (\ref{covLapF}) and the
Bianchi identity in (\ref{Bianchi1}) gives
\begin{eqnarray*}
\nabla_{\alpha} \nabla^{\alpha} \sigma_{\mu}
= -2 \xi^{\delta} \F^{\alpha\beta} \C_{\alpha\beta\delta\mu}
+ \frac{2}{3} R \sigma_{\mu} - 4 \xi^{\delta} R^{\beta}_{\,\,\delta} \F_{\beta\mu}
- \sigma_{\beta} R^{\beta}_{\,\,\mu} 
+ 2 \nabla_{\mu} \left ( \xi^{\rho} \xi^{\sigma} R_{\rho\sigma} \right ) + \\
\hspace{5cm} + 2 i \eta_{\gamma\mu\alpha\beta} \nabla^{\alpha} \left (
\xi^{\beta} \xi^{\sigma} R^{\gamma}_{\,\,\,\sigma} \right ).
\end{eqnarray*}
Combining this expression with the
Weitzenb\"ock formula (\ref{Weiz}), the following identity for the
gradient of $\FF^2$ is obtained
\begin{eqnarray*}
\nabla_{\mu} \Fsq = 2 \xi^{\nu} \F^{\alpha\beta} \C_{\alpha\beta\nu\mu}
- \frac{2}{3} R \sigma_{\mu} + 4 \xi^{\nu} R^{\beta}_{\,\,\nu} \F_{\beta\mu}
+ 2 \sigma_{\beta} R^{\beta}_{\,\,\mu}.
\end{eqnarray*}

\section{Generalized Ehlers group.}

The standard definition of the Ehlers group \cite{Ehlers}, \cite{Geroch}
is as follows. Consider a strictly stationary\footnote{The Ehlers group is
defined similarly when the Killing vector is spacelike everywhere.}
vacuum spacetime $(\V,g)$ 
(i.e. a vacuum spacetime admitting a Killing vector which is timelike
everywhere). Take the quotient set
$\V / \vec{\xi}$ with respect to the orbits of the Killing vector. This
is locally a manifold (i.e.
given any point $p \in \V$,
there exists an open neighborhood $U_p$ of $p$ such that
$\N_p \equiv U_{p} / \vec{\xi}$ is a manifold). There exists a
well-known one-to-one
correspondence between tensor fields in $\N_p$ and tensor fields in $U_p$
which are Lie-constant along $\vec{\xi}$ and which are completely orthogonal
to $\vec{\xi}$. The symmetric tensor $\lambda g_{\alpha\beta}+ \xi_{\alpha}
\xi_{\beta}$ 
has these  properties and therefore defines 
a symmetric tensor $\gamma_{ij}$ in $\N_p$ which endows this space
with a Riemannian structure. The Ernst one-form $\sigma_{\mu}$
associated to  $\vec{\xi}$ is closed (see (\ref{dsigma}))
and hence exact in a suitably
chosen $U_p$. Let $\sigma$ be a complex scalar in $U_p$
satisfying $\sigma_{\mu} = \nabla_{\mu} \sigma$. This function
$\sigma$ defines a complex
scalar in the quotient $\N_p$. As shown by Geroch \cite{Geroch},
the knowledge of
$\N_p$, $\gamma_{ij}$ and $\sigma$ is sufficient to reconstruct 
locally the original spacetime $\V$.
The action of the Ehlers group is defined by 
leaving $\gamma_{ij}$ invariant and transforming $\sigma$ according
to the M\"obius map $\ssig = (\alpha \sigma + i \beta )/(i \gamma
\sigma + \delta)$,
where $\alpha, \beta, \gamma, \delta$
are arbitrary real constants satisfying $\alpha\delta + \beta \gamma= 1$.
As shown in \cite{Geroch}, the transformed spacetime is also a
solution of the Einstein vacuum field equations.
The group structure of
the Ehlers group can be obtained by applying first a transformation
$\ssig = (\alpha \sigma + i \beta )/(i \gamma \sigma + \delta)$, and
then a second transformation 
$\sssig = (\alpha' \ssig + i \beta' )/(i \gamma' \ssig + \delta')$ (where
$\alpha' \delta' + \beta' \gamma' = 1$). The result is
$\sssig = (\alpha'' \sigma + i \beta '' )/(i \gamma'' \sigma + \delta'')$
with  $\alpha'', \beta'', \gamma'', \delta''$ given by 
\begin{eqnarray}
\left ( \begin{array}{cc}
           \alpha'' & i \beta'' \\
           i \gamma'' & \delta'' 
           \end{array} 
\right ) =
\left ( \begin{array}{cc}
           \alpha' & i \beta' \\
           i \gamma' & \delta' 
           \end{array} 
\right ) 
\left ( \begin{array}{cc}
           \alpha & i \beta \\
           i \gamma & \delta 
           \end{array} 
\right ).
\label{multipl}
\end{eqnarray}
This expression shows, in particular, that the Ehlers group is isomorphic
to $SL(2, \Bbb{R})$. 

As discussed in the introduction, it is desirable to have a description of
the Ehlers group solely in terms of spacetime objects, i.e. without
passing through the manifold of trajectories. We shall
start by defining a much larger group of transformations which will
turn out to contain the Ehlers group as a particular case.
This group of transformations
is defined for an arbitrary $n$-dimensional manifold and it
describes the fundamental underlying structure of the Ehlers group.

Let $\U$ be an arbitrary $n$-dimensional smooth manifold and
let us denote by $\CH(\U)$ the algebra of
vector fields and by $\Lam (\U)$ the set of smooth one-form fields on $\U$. 
Let us define $G \subset \CH (\U) \times \Lam  (\U) $ by
$G = \left \{ (\vec{\zeta}, \bm{W}) \in \CH (\U) \times \Lam (\U )
 \,\,\, ;  \,\,\, \zeta^{\alpha} W_{\alpha}|_p > -1,  
\,\,\, \forall p \in \U \,\,\, \right \}$. 
We  define also $\Gzeta \equiv \{ \bm{W} \in \Lam (\U) \, \, \, ; \left (\vec{\zeta},
\bm{W} ) \in G \right \}$. Let $\ji$ denote
the set of smooth, symmetric, two-covariant tensor fields in $\U$.
The following map defines an action of $G$ on $\ji$
\begin{eqnarray}
T \, \, :  \, \, G \times \ji & \longrightarrow & \ji \nonumber\\
\left (\zeta^{\alpha}, W_{\beta}, g_{\alpha\beta}  \right ) & \longrightarrow &
T(\vec{\zeta}, \bm{W}, g )_{\alpha\beta}
  \equiv  \nn^2 g_{\alpha\beta}
 - \zeta_{\alpha} W_{\beta} - \zeta_{\beta} W_{\alpha} - \frac{\lambda}{\nn^2}
W_{\alpha} W_{\beta},
\label{transmet}
\end{eqnarray}
where $\zeta_{\alpha} \equiv g_{\alpha\beta} \zeta^{\beta}$, $\nn^2 \equiv
\zeta^{\alpha} W_{\alpha} + 1$ and $\lambda \equiv - \zeta^{\alpha} \zeta^{\beta}
g_{\alpha\beta}$. Given $(\vec{\zeta}, \bm{W} ) \in G$, we can
also define the map 
\begin{eqnarray}
\TzeW \, \, :  \, \,  \ji & \longrightarrow & \ji \\
g  & \longrightarrow &
\TzeW (g)  \equiv  T(\vec{\zeta}, \bm{W}, g).
\end{eqnarray}
In order to show that $\TzeW$ defines a group structure on $\Gzeta$
we need to compose two such transformations.
After a calculation we find
\begin{eqnarray*}
\Tzeta_{\sm{W^2}} \circ \Tzeta_{\sm{W^1}} 
=  \Tzeta_{\sm{W^2} + \nn_2^2 \sm{W^1}}, \hspace{1cm}
\mbox{where} \hspace{5mm} \nn^2_2 = 1 + 
\zeta^{\alpha} W^{2}_{\alpha}.
\end{eqnarray*}
It only remains to check that $\bm{W^2} + \nn_2^2 \bm{W^1} \in \Gzeta$,
which follows immediately from
\begin{eqnarray}
\zeta^{\alpha} \left ( W^{2}_{\alpha} + \nn^2_2 W^1_{\alpha} \right ) + 1 =
\nn_1^2 \nn_2^2 > 0,
\label{dnns}
\end{eqnarray}
where $\nn_1^2 = \zeta^{\alpha} W^1_{\alpha} + 1$. Thus, the composition
law in $\Gzeta$ induced by $\TzeW$ is 
\begin{eqnarray*}
\cdot  \, : \, \, \, \Gzeta \times \Gzeta & \longrightarrow & \Gzeta \\
     \left ( \bm{W^2} , \bm{W^1} \right )  & \longrightarrow & 
\bm{W^2} \cdot \bm{W^1} = \bm{W^2} + \left (1 + \zeta^{\alpha} W^{2}_{\alpha} \right ) \bm{W^1}.
\end{eqnarray*}
The unit element of $(\Gzeta, \cdot)$ is obviously $\bm{W} = 0$,
and the inverse of $\bm{W} \in \Gzeta$ is 
$ - \bm{W} / ( 1+ \zeta^{\alpha} W_{\alpha} )$. In order to investigate the
group structure further, let us find a suitable
set of one-parameter subgroups of $(\Gzeta,\cdot )$. 
Let  $\bm{W^0}$ be an arbitrary smooth one-form on $\U$ (not necessarily
satisfying $\zeta^{\alpha} W^{0}_{\alpha} +1 > 0$) and let us define
the smooth real function 
\begin{eqnarray*}
f_{\sm{W^0}} (t) = \left \{ \begin{array}{cl}
                       \frac{\exp {(\zeta^{\alpha} W^{0}_{\alpha} t)} -1  }{
                        \zeta^{\alpha} W^{0}_{\alpha}} &
                       \mbox{ at points where \hspace{2mm} }
                       \zeta^{\alpha} W^{0}_{\alpha} \neq 0 \\
                       t  & \mbox{ at points where \hspace{2mm} } 
                       \zeta^{\alpha} W^{0}_{\alpha} = 0
                       \end{array}
\right . .
\end{eqnarray*}
Thus, $\bm{W} (t) \equiv
f_{\sm{W_{0}}} (t) \bm{W^{0}}$ is a smooth one-form in $\U$.
It is a simple exercise to check that $\bm{W}(t)$ defines a
one-parameter subgroup of $\Gzeta$, i.e. that $\bm{W}(t) \in 
\Gzeta$ and that $\bm{W}(s) \cdot \bm{W}(t) = \bm{W}(t+s)$ for all
$t, s \in \Bbb{R}$. Since $\Lam(\U)$ is infinite dimensional, it follows that
$(\Gzeta, \cdot)$ is an infinite dimensional group 
which acts on the space of symmetric 2- covariant tensors. 

For definiteness, we shall now restrict ourselves to the case
in which $\U$ is four-dimensional. Our next aim is to show that
$\Gzeta$ leaves the subset of Lorentzian metrics on $\U$ invariant,
i.e. that the signature is preserved under the action of
$T^{\vec{\zeta}}_{\sm{W}}$.
\begin{lemma}
\label{group}
Let $\U$ be a four-dimensional manifold, $g$ be a Lorentzian metric
on $\U$ and $\bm{W} \in \Gzeta$. Then
$\gzeta$ is also a Lorentzian metric in $\U$.
Furthermore, if $(\U,g)$ is orientable with volume form
$\eta_{\alpha\beta\mu\nu}$, then $(\U,\gzeta)$ is also orientable
with volume form 
$\eeta_{\alpha\beta\gamma\delta} = \nn^2 \eta_{\alpha\beta\gamma\delta}$.
In addition, the inverse metric of $\gzeta_{\alpha\beta}$ reads 
\begin{eqnarray}
(\gzeta^{-1})^{\alpha\beta} = \frac{1}{\Omega^2} \left (
g^{\alpha\beta}  + W^2 \zeta^{\alpha} \zeta^{\beta} +
\zeta^{\alpha} W^{\beta} + W^{\alpha} \zeta^{\beta}
\right ),
\label{inver}
\end{eqnarray}
where $W^{\alpha} \equiv g^{\alpha\beta} W_{\beta}$, 
$W^2 \equiv W^{\alpha} W_{\alpha}$ and
$g^{\alpha\beta}$ is the inverse of $g_{\alpha\beta}$.
\end{lemma}
{\it Proof.}
We first notice the following identity, valid
in $n$ dimensions, 
\begin{eqnarray}
\det ( a_1 M_{ij} + a_2 b_{i} b_{j} + a_3 c_{i} c_{j} )
= a_1^{n-2} \det (M_{ij}) \left [ (a_1 + a_2 b^2) (a_1 + a_3 c^2) - a_2 a_3
(b  c)^2 \right ],
\label{iden}
\end{eqnarray}
where $a_1 \neq 0$, $a_2$, $a_3$ are arbitrary constants,
$M_{ij}$ is an arbitrary $n \times n$ invertible
matrix, $b_{i}$, $c_{i}$ are arbitrary $n$-column vectors and $b^2=
(M^{-1})^{ij} b_{i} b_{j}$, $c^2= (M^{-1})^{ij} c_{i} c_{j}$ and
$b  c = (M^{-1})^{ij} b_{i}
c_{j}$. This identity can be proven 
straightforwardly by first showing its validity when $a_3 =0$ (which
follows directly from the definition of determinant) and then applying
the result twice. We want to use (\ref{iden})
to evaluate the determinant
of (\ref{transmet}) in an arbitrary local coordinate system. At
points $p \in \U$ where $\vec{\zeta}$ is non-null with respect to
$g_{\alpha\beta}$ (i.e. $\lambda \neq 0$)  the tensor $\gzeta$ can be
rewritten as
\begin{eqnarray}
\gzeta_{\alpha\beta} = \nn^2 \left [ g_{\alpha\beta} + \frac{1}{\lambda}
 \left ( \zeta_{\alpha} \zeta_{\beta}-
V_{\alpha} V_{\beta} \right ) \right ]
\label{transmet2}
\end{eqnarray}
where $V_{\alpha} \equiv \zeta_{\alpha} + \frac{\lambda}{\nn^2} W_{\alpha}$. 
Applying (\ref{iden}) to this expression, we easily find 
 $\det{(\gzeta)}= \nn^4
\det (g)$. Similarly, at those points where $\vec{\zeta}$ is null (i.e.
$\lambda =0$), $\gzeta_{\alpha\beta}$ can be rewritten as
\begin{eqnarray*}
\gzeta_{\alpha\beta} = \nn^2 g_{\alpha\beta} - \frac{1}{2}
 \left ( \zeta_{\alpha} + W_{\alpha} 
\right ) \left  (\zeta_{\beta} + W_{\beta} \right ) + \frac{1}{2} 
\left ( \zeta_{\alpha} - W_{\alpha} \right ) \left  (\zeta_{\beta} -
 W_{\beta} \right ).
\end{eqnarray*}
Applying the identity (\ref{iden}) to this expression we obtain again
$\det{(\gzeta)}= \nn^4 \det (g)$.
Thus, $\gzeta$ is invertible everywhere. Expression (\ref{inver}) for
the inverse metric can be checked by simple calculation.
So, it only remains to show that the signature of 
$\gzeta_{\alpha\beta}$ is $(-1,1,1,1)$. This is proven by noticing that
any element $\bm{W} \in \Gzeta$ can be continuously connected
to the identity. Indeed, let $\bm{W}$ be an arbitrary one-form
satisfying $\zeta^{\alpha} W_{\alpha} + 1 > 0$ everywhere. Define
$\bm{W^0} = \bm{W}$ at points where $\zeta^{\alpha} W_{\alpha} = 0$ and
$\bm{W^0} = \left ( \zeta^{\alpha} W_{\alpha} \right )^{-1} \ln
\left [ 1 + \zeta^{\beta} W_{\beta} \right ]  \bm{W}$ at points where
$\zeta^{\alpha} W_{\alpha} \neq 0$. It follows that $\bm{W^0}$ is a smooth
one-form and therefore we can define its associated one-parameter
subgroup $\bm{W}(t)$ according to
the procedure above. It is easy to check that $\bm{W}(1) = \bm{W}$.
Since $ \det (\Tzeta_{\sm{W}(t)} (g))$ is non-zero everywhere for all $t$, 
it follows that the signature
must remain unchanged. $\hfill \Box$

The group $\Gzeta$ is defined for any manifold and any smooth 
vector field $\vec{\zeta}$. It is not even necessary that $\U$ is a 
Riemannian space. For an $n$-dimensional $\U$, the transformation
defined by $G$ contains $2n-1$ arbitrary functions ($2n$ functions 
are necessary to define $\vec{\zeta}$ and $\bm{W}$ but the transformation
(\ref{transmet}) has the explicit symmetry
$T( K \vec{\zeta}, K^{-1} \bm{W}, g) =
T( \vec{\zeta}, \bm{W}, g )$ where $K$ is a nowhere vanishing scalar function).
So, $G$ defines a very general transformation. Obviously,
vacuum solutions are not mapped into vacuum solutions in general.
However, as we shall see below, $G$ contains the Ehlers group as a particular
case. Thus, it is plausible that there may exist other
interesting subsets of $G$. Exploring this problem is beyond the scope
of this paper and should be addressed elsewhere.

The aim of the next section is to exploit the general group structure
introduced in this section in order to define the Ehlers group on spacetimes
admitting a Killing vector of arbitrary causal character.

\section{Spacetime description of the Ehlers group.}

Let us consider a spacetime $(\M,g)$ admitting a Killing
vector $\vec{\xi}$ such that $\xi^{\alpha} R_{\alpha\beta}=0$ and let
us define all the objects associated to $\vec{\xi}$ as in section 2.
The Ernst one-form is closed by virtue of (\ref{dsigma}).
In order to define the Ehlers group 
we need to impose two global requirements on $\M$, which are 
essential for the whole construction.
The first one is that $\bm{\sigma}$ is exact, i.e. that there exists a
complex function $\sigma \equiv \lambda - i \omega$ on $\M$ 
($\omega$ is called twist potential),
such that $\nabla_{\alpha} \sigma = \sigma_{\alpha}$. 
Under these circumstances, let us define the 2-form
\begin{eqnarray}
4 \gamma \left [ \delta F^{\star}_{\alpha\beta} + \gamma \left (
\omega  F^{\star}_{\alpha\beta} - \lambda F_{\alpha\beta} \right ) \right ]
= \mbox{Re} \left [- 4 \gamma \left (\gamma \ov{\sigma} + i \delta \right )
\F_{\alpha\beta} \right ], 
\label{twoform}
\end{eqnarray}
where $\gamma$ and $\delta$ are arbitrary, non-simultaneously vanishing,
real constants (a bar denotes complex conjugate). The following chain
of equalities shows that this 2-form
is closed,
\begin{eqnarray}
\star \bm{d} \left ( 
\mbox{Re} \left [ \left ( \gamma \ov{\sigma} + i \delta \right )
\FF \right ] \right ) =
\mbox{Re} \left [ \star \bm{d} \left (
 \left ( \gamma \ov{\sigma} + i \delta \right )
\FF \right ) \right ] =
\gamma \mbox{Re} \left [ \star \left ( \FF \wedge \ov{\bm{\sigma}} \right )
\right ]  = \nonumber \\
\hspace{7cm} 
\gamma \mbox{Re} \left ( \mbox{i}_{\ov{\bm{\sigma}}} \star \FF \right ) = 
\gamma \mbox{Re} \left ( - i \, \mbox{i}_{\ov{\bm{\sigma}}} \FF \right ) = 0.
\label{WW2}
\end{eqnarray}
In the second equality we used the fact that $\FF$ is closed 
(see (\ref{cod})). In the
third equality we used the general identity $\star 
( \bm{\alpha} \wedge \bm{\beta} ) = \mbox{i}_{\sm{\beta}} \star \bm{\alpha}$
where $\bm{\alpha}$ is any $p-$form, and $\bm{\beta}$ is any $q-$form 
($q+p \leq \mbox{dim} \, \M$) and
$\mbox{i}_{\sm{\beta}} \bm{\alpha}$ denotes interior contraction.
The last
equality is a consequence of $\F_{\nu\mu} \ov{\F}_{\delta}^{\,\,\,\,\mu}$
being symmetric and therefore real.

Now we have to impose the second
global requirement on  $(\M,g)$, namely that the 2-form
(\ref{twoform}) is
exact for all values of $\gamma$ and $\delta$,
i.e. there exists a one-form $\bm{W}$ satisfying
\begin{eqnarray}
\nabla_{\alpha} W_{\beta} - \nabla_{\beta} W_{\alpha} 
= \mbox{Re} \left [- 4 \gamma \left (\gamma \ov{\sigma} + i \delta \right )
\F_{\alpha\beta} \right ]. 
\label{WW}
\end{eqnarray}
Furthermore, we demand that there exists a solution $\bm{W}$ of
(\ref{WW}) that satisfies
\begin{eqnarray}
\nn^2 \equiv  \xi^{\alpha} W_{\alpha} + 1 = 
\gamma^2 \lambda^2 + \left ( \delta + \gamma \omega \right )^2 
= \left ( i \gamma \sigma + \delta \right )
\left ( - i \gamma \ov{\sigma} +\delta \right ).
\label{nn2}
\end{eqnarray}
Since the solution of (\ref{WW}) is defined up to a closed one-form, we can
always achieve (\ref{nn2}) locally. However, it is essential that 
(\ref{nn2}) is satisfied also globally. In general, (\ref{WW}) and
(\ref{nn2}) do not fix a unique solution yet (we can always
add a closed one-form which is orthogonal to $\vec{\xi}$, for
instance $\bm{\sigma}$).
However, provided the global conditions discussed above are fulfilled,
we can associate to each pair of values $\gamma$ and
$\delta$ a unique one-form $\bm{W}$ satisfying (\ref{WW})-(\ref{nn2}).
We will assume from now on that such a choice has been made.

Now, we can apply the
general transformation $T^{\vec{\xi}}_{\sm{W}}$ defined in Section 3
with respect to the Killing vector $\vec{\xi}$ and the
one-form $\bm{W}$.
In order to do that, we must ensure  that
$\xi^{\alpha} W_{\alpha} +1 >0$.
Thus, the points where $\nn^2$ vanishes must be excluded
from $\M$ beforehand. These points correspond 
to $\lambda=0$ and $\omega = - \delta/\gamma$ (with
$\gamma \neq 0$). Notice that the
excluded points (if any) are  contained in the region
where  the Killing vector is null. As we shall see below,
the transformed spacetime will in general contain a curvature singularity
precisely at the points we have excluded (if they exist).
Thus, from now on, and for each value of $\delta/\gamma$, we exclude
the set of points $\lambda =0$ and
$\omega= - \delta / \gamma$ from the manifold $\M$. In order
to simplify the notation, the resulting manifold (which is in general
different for each value of $\delta/\gamma$) will still be denoted by
$\M$. The meaning of $\M$ should become clear from the context.

We can now define the transformed metric 
$\g_{\alpha\beta}$ on $\M$. From Section 3 we know
that $\g_{\alpha\beta}$ is smooth and Lorentzian. Our next aim
is to prove that $\vec{\xi}$ is also a Killing vector of
$\g_{\alpha\beta}$.
Then, we will prove that its Ricci tensor (which we denote by
$\ric_{\alpha\beta}$) 
satisfies $\xi^{\alpha} \ric_{\alpha\beta} =0$. This will allow us
to compose transformations and prove that they define a group.
\begin{lemma}
The vector field $\vec{\xi}$ is a Killing vector of the metric 
$\g_{\alpha\beta}$.
\end{lemma}
{\it Proof:} By construction $\pounds_{\vec{\xi}}$$\,\nn=0$.
Thus,  we only need to
show that $\pounds_{\xi} \bm{W}$$=0$.
From the definition of Lie derivative we find
$\left ( \pounds_{\vec{\xi}} \bm{W} \right )_{\alpha} = \xi^{\mu} \left (
\nabla_{\mu} W_{\alpha} - \nabla_{\alpha} W_{\mu} \right )
+ \nabla_{\alpha} \left (\mbox{$\nn^2$} -1 \right )$. Using the equation 
\begin{eqnarray}
\nabla_{\alpha} \nn^2 = 2 \gamma \mbox{Re} \left [
\left ( \gamma \ov{\sigma} + i \delta \right ) \sigma_{\alpha} \right ],
\label{nabnn}
\end{eqnarray}
(which is a direct consequence of (\ref{nn2}))
the vanishing of $\pounds_{\vec{\xi}} \bm{W}$ follows immediately.
$\hfill \Box$.

In order to be able to compose transformations  $\Txi_{\sm{W}}$,
it is necessary to determine the Killing form associated to $\vec{\xi}$ in
the transformed spacetime $\left (\M, \g \right)$.
This is addressed in the following Lemma.
\begin{lemma}
The complex two-form
\begin{eqnarray} 
\FFpr =  \frac{1}{\left (\delta + i \gamma \sigma  \right )^2}
 \left [ \nn^2  \FF - \frac{1}{2}
\bm{W} \wedge \bm{\sigma} \right ]
\label{Fpr1}
\end{eqnarray}
is the Killing form of $\vec{\xi}$ with
respect to the metric $\g_{\alpha\beta}$.
\end{lemma}
{\it Proof:}
We must show that $\FFpr$ is self-dual with respect to the metric
$\g_{\alpha\beta}$ and also that  $\mbox{Re} (\FFpr) =
\frac{1}{2}\bm{d} \bm{V}$ where $V_{\alpha} \equiv
\g_{\alpha\beta} \xi^{\beta}$. To prove the first part,
let us raise the indices of
$\FFpr_{\alpha\beta}$ with respect to $\g_{\alpha\beta}$. Starting from
\begin{eqnarray}
\Fpr_{\alpha\beta} =  \frac{1}{\left (\delta + i \gamma \sigma  \right )^2}
 \left [ \nn^2 \ \F_{\alpha\beta}
+ \frac{1}{2}
\left (W_{\beta} \sigma_{\alpha} - W_{\alpha} \sigma_{\beta} \right )
\right ],
\label{Fpr}
\end{eqnarray}
a somewhat long, although straightforward, calculation gives
\begin{eqnarray}
\Fpr^{\mu\nu} \equiv 
\left[\g^{-1} \right]^{\mu\alpha} 
\left [\g^{-1} \right ]^{\nu\beta} \Fpr_{\alpha\beta} =
\frac{1}{\nn^2 \left ( i \gamma \sigma + \delta \right )^2 }
\left [ \F^{\mu\nu} + \xi^{\mu} W_{\alpha} \F^{\alpha\nu} -
\xi^{\nu} W_{\alpha} \F^{\alpha\mu} \right ].
\label{Fprupup}
\end{eqnarray}
Using identity (\ref{xiF}) and the fact that $\eeta_{\alpha\beta\lambda\mu}
= \nn^2 \eta_{\alpha\beta\lambda\mu}$ (see Lemma \ref{group}),
the self-duality of 
$\Fpr_{\alpha\beta}$ follows readily. Regarding $\mbox{Re} (\FF) =
\frac{1}{2} \bm{d} \bm{V}$, we first recall that
$V_{\alpha} = \xi_{\alpha} + \frac{\lambda}{\nn^2}
W_{\alpha}$. Using equations (\ref{WW}) and (\ref{nabnn}) together 
with the fact that $\nabla_{\alpha} \lambda = \mbox{Re} (\sigma_{\alpha} )$,
the equation $\nabla_{\alpha} V_{\beta} - \nabla_{\beta} V_{\alpha} =
2 \, \mbox{Re} \, (\Fpr_{\alpha\beta})$ follows without difficulty.
$\hfill \Box$

Once we know the Killing form of $\vec{\xi}$ in the transformed spacetime, we
can compute the Ernst one-form of $\vec{\xi}$ with respect to
$\g_{\alpha\beta}$. From the definition (\ref{defsigma}) we find
\begin{eqnarray*}
\ssig_{\alpha} \equiv 2 \xi^{\beta} \Fpr_{\beta\alpha} = 
\frac{\sigma_{\alpha}}{\left ( i \gamma \sigma + \delta \right )^2}.
\end{eqnarray*}
This equation implies that $\ssig_{\alpha}$ is exact on $\M$, i.e.
$\ssig_{\alpha} = \nabla_{\alpha} \ssig$, where the function
$\ssig$ can be written in the following form, which is valid
for all possible values of $\gamma$ and $\delta$, 
\begin{eqnarray}
\ssig = \frac{\alpha \sigma + i \beta}{i \gamma \sigma + \delta},
\label{transformsigma}
\end{eqnarray}
where $\alpha, \beta$ 
are real constants satisfying $\alpha\delta + \beta \gamma= 1$.
 
The transformation law for $\Fsq$ will be required in section 6. It
is found directly from (\ref{Fprupup}) and reads
\begin{eqnarray}
\Fprsq = \frac{\Fsq}{\left (\delta + i \gamma \sigma \right )^4}.
\label{Fsqlaw}
\end{eqnarray}

Let us now prove that $\xi^{\alpha} \ric_{\alpha\beta}=0$. This is
important because it will allow us to compose
transformations (recall that the only local conditions we imposed on $(\M,g)$
in order to define $\T$ were the existence of a Killing vector
$\vec{\xi}$ and that $\xi^{\alpha} R_{\alpha\beta}=0$).
\begin{lemma}
The Ricci tensor of $\g_{\alpha\beta}$ satisfies
$\xi^{\alpha} \ric_{\alpha\beta} = 0$.
\end{lemma}
{\it Proof:}
From Lemma \ref{lemmaone}, $\xi^{\alpha} \ric_{\alpha\beta} = 0$
is equivalent to $\FFpr$ being closed.
Since $\bm{\sigma}$ is exact and 
$\bm{W}$ satisfies $\bm{d} \bm{W} 
= \mbox{Re} \left [- 4 \gamma \left (\gamma \ov{\sigma} + i \delta \right )
\FF \right ]$, we obtain, after making use of
the definition of $\nn^2$,
\begin{eqnarray*}
\bm{d} \FFpr = \frac{-i \gamma}{\delta + i \gamma \sigma} \ov{\bm{\sigma}}
\wedge \FF - \frac{i \gamma \left (\delta - i \gamma \ov{\sigma} 
\right )}{\left ( \delta + i \gamma \sigma \right )^2} \bm{\sigma}
\wedge \FF + \frac{2 \gamma}{\left (\delta + i \gamma \sigma \right )^2}
\mbox{Re} \left [ \left (\gamma \ov{\sigma} + i \delta \right ) \FF \right ]
\wedge \bm{\sigma} = 0.
\end{eqnarray*}
The vanishing of this expression follows from expanding the real part
of the third term and using the identity 
$\FF \wedge \ov{\bm{\sigma}} + \ov{\FF} \wedge \bm{\sigma} = 0$, which has 
already been proven in (\ref{WW2}).
$\hfill \Box$

Thus, the transformations defined by $\bm{W}$ can be
iterated, at least locally. However, we also
had to impose two global conditions on $(\M,g)$
in order to define $\T$. The first one (i.e. that the 
Ernst one-form is exact) has already been proven in
(\ref{transformsigma})). Regarding $\bm{W}$,
we must ensure that the
exterior system (\ref{WW}) in the transformed spacetime is also integrable.
We address this issue as follows. Consider the transformed equation
\begin{eqnarray}
\nabla_{\alpha} W^{\prime}_{\beta} - \nabla_{\beta} W^{\prime}_{\alpha} 
= \mbox{Re} \left [- 4 \gamma^{\prime} \left (\gamma^{\prime}
\, \ov{\ssig} + i \delta^{\prime} \right )
\Fpr_{\alpha\beta} \right ], 
\label{WWpr}
\end{eqnarray}
where $\gamma^{\prime}$ and $\delta^{\prime}$ are not simultaneously zero.
The left-hand side of (\ref{WWpr}) is already known to be closed, so we only 
need to prove that it is also exact.

Let us assume for the
moment that $\bm{W^{\prime}}$  defined by (\ref{WWpr})
exists globally and let us use the
results in Sect. 3 in order to find its
explicit expression.
We first apply the transformation corresponding to 
$(\alpha,\beta,\gamma,\delta)$ to the original metric $g_{\alpha\beta}$
and construct  $\g_{\alpha\beta}$. Then, we apply the 
transformation with respect to $\bm{W^{\prime}}$ to
obtain a second metric $T^{\vec{\xi}}_{\sm{W'}}\left (\g
\right)_{\alpha\beta}$. From the results in Sect. 3 we know that
$T^{\vec{\xi}}_{\sm{W'}}\left (\g
\right)_{\alpha\beta} = 
T^{\vec{\xi}}_{\sm{W''}} (g)_{\alpha\beta}$, where 
$\bm{W''} = \bm{W'} + {\nn'}^2 \bm{W}$, and ${\nn'}^2 =
1 + \xi^{\alpha} W^{\prime}_{\alpha} =
\left ( \delta ' + i \gamma ' \ssig \right) \left ( \delta' -
i \gamma ' \ssig  \right)$. If it were true that the transformations
defined by $\bm{W}$ form a group  isomorphic to
$SL(2, \Bbb{R})$, then $\bm{W''}$ should be the solution
of (\ref{WW}) corresponding to $\alpha'', \beta'', \gamma'', \delta''$,
(where these constants are given by (\ref{multipl})).
With this information at hand, we can
now prove that $\bm{W^{\prime}}$ exists and has the desired properties.

Let $\bm{W}$ and $\bm{W''}$ be the unique solutions 
of (\ref{WW}) corresponding to $\alpha, \beta, \gamma, \delta$
and $\alpha'', \beta'', \gamma'',\delta''$ respectively (they exist globally
on $\M$ by assumption). Let us define 
$\alpha', \beta', \gamma', \delta'$ as the solution of 
(\ref{multipl}) (which is unique). 
We can also define $\ssig$ according
to (\ref{transformsigma}) and
${\nn'}^2 \equiv
\left ( \delta ' + i \gamma ' \ssig \right) \left ( \delta' -
i \gamma ' \ssig  \right)$. Finally we define
\begin{eqnarray}
\bm{W'} \equiv \bm{W''} - {\nn'}^2 \bm{W}.
\label{newW}
\end{eqnarray}
This object exists globally on $\M$ by construction and we have shown that
it is the only
candidate for being the solution of (\ref{WWpr}) satisfying
$\xi^{\alpha} W'_{\alpha} +1 = {\nn'}^2$ (i.e. the only candidate
compatible with the existence of an isomorphism between the group we are constructing
and $SL(2,\Bbb{R})$).
We need to prove that $\bm{W'}$ thus defined actually 
solves the desired equations.
Checking $\xi^{\alpha} W'_{\alpha} +1 = {\nn'}^2$
is easy from
\begin{eqnarray*}
1 + \xi^{\alpha} W'_{\alpha}  = 1 + \xi^{\alpha} W''_{\alpha}
- {\nn'}^2 \xi^{\alpha} W_{\alpha} = {\nn'}^2 + {\nn''}^2 - {\nn'}^2 \nn^2, 
\end{eqnarray*}
after taking into account the definitions of $\nn$, ${\nn''}$ and ${\nn'}$
together with $i \gamma'' \sigma + \delta'' = \left(
i \gamma' \ssig + \delta' \right ) \left (i \gamma \sigma + \delta
\right)$. Finally, we need to check that  the differential equation
\begin{eqnarray*}
\bm{d} \bm{W'} = \bm{d} \left ( \bm{W''} - {\nn'}^2 \bm{W} \right ) 
= \mbox{Re} \left [- 4 \gamma' \left (\gamma' \,\ov{\ssig} +
i \delta' \right )
\FFpr\right ]
\end{eqnarray*}
is satisfied. This can be proven by direct calculation using 
the differential equations satisfied by $\bm{W''}$ and $\bm{W'}$ together
with the expression for
$\FFpr$ (\ref{Fpr1}). 

This argument proves that 
the composition of transformations can now be globally performed and that 
they forms a group isomorphic to $SL(2,\Bbb{R})$.
It should be remarked  that $\bm{W}$ can be chosen freely 
for each value of $\alpha, \beta, \gamma, \delta$
{\it only} for the original metric $g_{\alpha\beta}$. The corresponding
choice for the transformed metrics $\g_{\alpha\beta}$ is uniquely fixed by
(\ref{newW}).

It is easy to check that the group we have just defined
coincides with the original Ehlers group when 
the Killing vector is timelike (or spacelike).
Indeed, this follows from 
the transformation law for the metric when written as in (\ref{transmet2}),
and from the transformation law for the Ernst potential 
(\ref{transformsigma}). In the approach we have followed no restriction on the
causal character of the Killing vector was made. This shows that the Ehlers
group exists even when the Killing vector changes its causal character
throughout the spacetime. Furthermore, the global conditions necessary 
for the transformation to exist on a given spacetime have been clarified.
This extended group of transformations will be called {\it spacetime Ehlers
group} in this paper.

Although the spacetime Ehlers group has been defined for smooth spacetimes 
admitting a Killing vector $\vec{\xi}$ which satisfies
$\xi^{\alpha} R_{\alpha\beta}=0$, the vacuum subcase is
particularly important because the original Ehlers group maps vacuum
solutions into vacuum solutions. Proving that this also holds
for the  spacetime Ehlers group will be our next aim. There
are several methods to do this. We give a proof that exploits the group
structure and which might be of interest for 
more general situations (i.e. for other subsets of $G^{\vec{\xi}}$).

Let us consider an arbitrary one-parameter subgroup
of the spacetime Ehlers group
(i.e. of $SL(2,\Bbb{R})$). The starting metric
$g_{\alpha\beta}$ is transformed under this subgroup into
a one-parameter family of metrics
$\Txi_{\sm{W}(t)}(g)$, all of which satisfy $\xi^{\alpha}
R_{\alpha\beta}(t)=0$. Let us now assume that $g$ is vacuum. The group
structure of the spacetime Ehlers group implies
that the one-parameter family of metrics $\Txi_{\sm{W}(t)}(g)$ is vacuum for
all $t$ if and only if the linearized Einstein equations around the metric
$\Txi_{\sm{W}(t)} (g)$ are solved by the symmetric tensor
$\frac{d}{dt} \Txi_{\sm{W}(t)}(g)$. Let us recall that the linearized Einstein
vacuum field equations for a perturbation $h_{\alpha\beta}$
around a given metric $g_{\alpha\beta}$ are
\begin{eqnarray}
\dot{R}_{\alpha\gamma} \equiv
 - \frac{1}{2} \nabla_{\alpha} \nabla_{\gamma} \HH
- \frac{1}{2} \nabla^{\beta} \nabla_{\beta} \HH_{\alpha\gamma}
+ \frac{1}{2} \nabla^{\beta} \nabla_{\gamma} \HH_{\alpha\beta}
+ \frac{1}{2} \nabla^{\beta} \nabla_{\alpha} \HH_{\gamma\beta} = 0,
\label{linear}
\end{eqnarray}
where $ \HH = \HH_{\alpha\beta} g^{\alpha\beta}$. The following Lemma
is used in Theorem \ref{vac} below 
and gives the form of the linearized equations for the case
under consideration.

\begin{lemma}
\label{algebra}
Let $(\M,g)$ be a smooth spacetime which admits a Killing vector $\vec{\xi}$
and satisfies $\xi^{\alpha} R_{\alpha\beta}= 0$.
Let $\bm{W^0}$ be an arbitrary one-form which is Lie-constant along 
$\vec{\xi}$. Then, the linearized Einstein field
equations (\ref{linear}) for a symmetric tensor of the form
${\HH}_{\alpha\beta} = (\xi^{\nu}W^0_{\nu}) g_{\alpha\beta}
- \xi_{\alpha} W^0_{\beta} - \xi_{\beta} W^0_{\alpha}$ read
\begin{eqnarray}
-2 \dot{R}_{\alpha\beta} =
g_{\alpha\beta} \nabla_{\gamma} \nabla^{\gamma}
\left ( \xi^{\nu}W^0_{\nu} \right ) + 2 F_{\gamma\alpha} K_{\beta}^{\,\,\,
\gamma} + 2 F_{\gamma\beta} K_{\alpha}^{\,\,\,\gamma} + \xi_{\alpha} 
\nabla^{\gamma} K_{\beta\gamma} + \xi_{\beta} \nabla^{\gamma} K_{\alpha\gamma}
= 0, \label{fin}
\end{eqnarray}
where $F_{\alpha\beta} = \nabla_{\alpha} \xi_{\beta}$ and
$K_{\alpha\beta} = \nabla_{\alpha} W^0_{\beta} - \nabla_{\beta}
W^0_{\alpha}$.
Furthermore, if $W^0_{\alpha}$ satisfies the equations
\begin{eqnarray}
\nabla_{\alpha} W^0_{\beta} - \nabla_{\beta} W^0_{\alpha} = 4 a_1
F^{\star}_{\alpha\beta}, \hspace{2cm} \xi^{\alpha}
 W^0_{\alpha} = 2 \left (a_2 + a_1 \omega \right ),
\label{eqW2}
\end{eqnarray}
where $a_1$ and $a_2$ are arbitrary constants and $\omega$ is the twist
potential of $\vec{\xi}$, then the linearized equations (\ref{fin})
are identically satisfied.
\end{lemma}
{\it Proof:}
Inserting the expression for ${\HH}_{\alpha\beta}$ into (\ref{linear}) and
expanding the appropriate second covariant derivatives we obtain,
after using the
Killing equations for $\vec{\xi}$ and $\nabla^{\mu}
\nabla_{\alpha} \xi_{\mu} = \xi^{\mu} R_{\mu\alpha}= 0$,
\begin{eqnarray}
-2 \dot{R}_{\alpha\beta}= 
2 \nabla_{\alpha} \xi_{\gamma} \nabla^{\gamma} W^0_{\beta} + 2 \nabla_{\beta}
\xi_{\gamma} \nabla^{\gamma} W^0_{\alpha} + \xi^{\gamma}
\nabla_{\gamma} \left (
\nabla_{\beta} W^0_{\alpha} + \nabla_{\alpha} W^0_{\beta} \right ) 
\nonumber \\
+ g_{\alpha\beta}
\nabla^{\gamma} \nabla_{\gamma} \left (\xi^{\nu} W^0_{\nu} \right ) 
+ \nabla^{\gamma} \left [ \xi_{\alpha} \left (\nabla_{\beta} W^0_{\gamma}
- \nabla_{\gamma} W^0_{\alpha} \right ) + \xi_{\beta} \left (
\nabla_{\alpha} W^0_{\gamma} - \nabla_{\gamma} W^0_{\alpha} \right ) \right ].
\label{inter}
\end{eqnarray}
Transforming this expression into (\ref{fin}) is not difficult after
using $\pounds_{\vec{\xi}}\,$$\,(\nabla_{\alpha} W^0_{\beta})=0$,
(which is a direct consequence of
$\bm{W^0}$ being Lie-constant along $\vec{\xi}$) in the third term
of  (\ref{inter}). 

To prove the second part of the Lemma,
we first notice that equation (\ref{eqW2}) is locally
integrable because
$\bm{F^{\star}}$ is closed (from $\nabla^{\mu} F_{\mu\nu} =
\xi^{\mu} R_{\mu\nu}= 0$). Furthermore
\begin{eqnarray*}
\nabla^{\beta} F^{\star}_{\gamma\beta} = \frac{1}{2}
\eta_{\gamma\beta\rho\sigma} \nabla^{\beta} \nabla^{\rho} \xi^{\sigma} =
\frac{1}{2} \eta_{\gamma\beta\rho\sigma} \xi^{\mu} R_{\mu}^{\,\,\,
\beta\rho\sigma} = 0,
\end{eqnarray*}
by virtue of the first Bianchi identities. Moreover, the imaginary part
of the first identity in (\ref{chichi}) applied
to $\cm{X} = \cm{Y} = \FF$ and the imaginary part
of identity (\ref{divsigma}) read
\begin{eqnarray*}
\nabla_{\beta} \nabla^{\beta} \omega = 2 F_{\alpha\beta} F^{\star\alpha\beta},
\hspace{1cm}
F_{\beta\alpha} F^{\star\beta}_{\,\,\,\,\,\,\gamma} 
+F_{\beta\gamma} F^{\star\beta}_{\,\,\,\,\,\,\alpha} = \frac{1}{2} g_{\alpha\gamma}
F_{\rho\sigma} F^{\star\rho\sigma}.
\end{eqnarray*}
Using these expressions, the vanishing of (\ref{fin}) follows
straightforwardly. $\hfill \Box$

We can now prove the following theorem which shows that the transformed metric of a vacuum
metric is also vacuum. In this theorem, all the
global conditions required for the existence of the Ehlers transformation are
spelled out.
\begin{theorem}
\label{vac}
Let $(\M,g)$ be a smooth spacetime admitting a Killing vector $\vec{\xi}$ and
satisfying Einstein vacuum field equations. Let $\delta, \gamma \in \Bbb{R}$ satisfy
$\delta^2 + \gamma^2 \neq 0$. Define $\lambda$,
$\FF$ and $\bm{\sigma}$ as the squared norm, the Killing form and the
Ernst one-form associated to $\vec{\xi}$. If the two following conditions are satisfied

1. The Ernst one-form is exact, i.e. it exists a complex smooth function $\sigma
\equiv \lambda - i \omega$ such that $\bm{\sigma} = \bm{d} \sigma$.

2. The closed two-form $\mbox{Re}
\left ( - 4 \gamma \left (\gamma \ov{\sigma} +  i \delta \right )
\FF \right )$ is exact and the equation ${\bm d} \bm{W} =$

\noindent $\mbox{Re} \left ( - 4 \gamma \left (\gamma \ov{\sigma} + 
 i \delta \right ) \FF \right )$
admits a solution satisfying $W_{\alpha} \xi^{\alpha} + 1 =
\left ( i \gamma \sigma + \delta \right ) \left ( - i \gamma \ov{\sigma} + \delta \right)
\equiv \Omega^2$.

Then, the symmetric tensor $\T (g) \equiv \Omega^2 g - \bm{\xi} \otimes \bm{W} -
\bm{W}\otimes \bm{\xi} - \frac{\lambda}{\Omega^2} \bm{W} \otimes \bm{W}$ defines a smooth vacuum
metric on the  spacetime $\tilde{\M} = \left \{ p \in \M ; \lambda |_p \neq 0 \mbox{ or }
(\gamma \omega + \delta) |_p \neq 0 \right \}$.
\end{theorem}
{\bf Remark.} As mentioned before, conditions 1. and 2. are always
fulfilled locally. So they only pose global obstructions to the existence
of the Ehlers group.

\vspace{3mm}

\noindent {\it Proof:}
From the group structure, we have
\begin{eqnarray}
\frac{d}{dt} \Txi_{\sm{W}(t)} (g) = \left . \frac{d}{ds}
\left [ \Txi_{\sm{W}(s)} \left ( \Txi_{\sm{W}(t)} (g) \right ) 
\right ] \right |_{s=0}.
\label{complaw}
\end{eqnarray}
On the other hand, the general transformation law (\ref{transmet}) implies
that, for an arbitrary symmetric tensor $\tilde{g}_{\alpha\beta}$, we have
\begin{eqnarray}
\left .\frac{d}{ds} \Txi_{\sm{W}(s)} \left (\tilde{g} \right )_{\alpha\beta}
\right |_{s=0}
= \left (\xi^{\mu}  \left .\frac{dW_{\mu} (s)}{ds} \right |_{s=0}
\right ) \tilde{g}_{\alpha\beta} - \tilde{\xi}_{\alpha} 
 \left .\frac{dW_{\beta} (s)}{ds} \right |_{s=0} -
\tilde{\xi}_{\beta}
 \left .\frac{dW_{\alpha} (s)}{ds} \right |_{s=0},
\label{linearizedG}
\end{eqnarray}
where $\tilde{\xi}_{\alpha} = \tilde{g}_{\alpha\beta} \xi^{\beta}$
and we used the fact that $\bm{W}(0)=0$. Combining 
(\ref{complaw}) and (\ref{linearizedG}) we get
\begin{eqnarray*}
\frac{d}{dt} \Txi_{\sm{W}(t)} (g)_{\alpha\beta} 
= \left (\xi^{\mu}  W^{0}_{\mu} \right ) \Txi_{\sm{W}(t)} (g)
_{\alpha\beta} - V(t)_{\alpha}  W^{0}_{\beta}
- V(t)_{\beta} W^0_{\alpha},
\end{eqnarray*}
where $V(t)_{\alpha} = \Txi_{\sm{W}(t)}(g)_{\alpha\beta} \xi^{\beta}$
and $\bm{W^0} \equiv \frac{d \sm{W} (s)}{ds} |_{s=0}$. The equations satisfied
by $\bm{W^{0}}$ 
can be obtained directly from (\ref{WW}) and (\ref{nn2}) after using
$\gamma(s)|_{s=0} = 0$ and $\delta(s) |_{s=0} = 1$ and read
\begin{eqnarray*}
\nabla_{\alpha} W^0_{\beta} - \nabla_{\beta} W^{0}_{\alpha}
= \mbox{Re} \left [ -4 i a_1 \F^{\prime}_{\alpha\beta} \right ]
= 4 a_1 F^{\prime \star}_{\alpha\beta}, \hspace{1cm}
\xi^{\alpha} W^0_{\alpha} = 2 \left (a_2 + a_1 w^{\prime} \right),
\end{eqnarray*}
where $\F^{\prime}_{\alpha\beta}$ and $\omega^{\prime}$ are the Killing form
and twist potential of $\vec{\xi}$ in the metric 
$\Txi_{\sm{W}(t)}(g)_{\alpha\beta}$ and we have
defined $a_1= \frac{d \gamma(s)}{ds} |_{s=0}$
and $a_2 = \frac{d \delta(s)}{ds} |_{s=0}$. Thus, all the conditions of
Lemma \ref{algebra} are fulfilled
and we can conclude that $\Txi_{\sm{W}(t)} (g)$ satisfies Einstein vacuum field equations.
The theorem follows after using that $SL(2,\Bbb{R})$ is connected and 
that any element of a connected Lie group
$G_0$ can be expressed as a finite product of elements in $\mbox{exp}
({\cal G})$ (where  ${\cal G}$ is the Lie algebra of $G_0$ and
$\mbox{exp}$ is the exponential
map) (see e.g. \cite{Corn}). $\hfill \Box$

\section{Action of the Ehlers group on the Weyl tensor.}
\label{SecWeyl}

Our aim in this section is to obtain an explicit expression for
the Weyl tensor of the Ehlers transformed spacetime. In order to do that 
we will exploit the identity (\ref{derivsigma}) of section 2. More
precisely, we want  to evaluate the left-hand side of 
this identity for the transformed metric $\g_{\alpha\beta}$ in order to
get an expression for the transformed Weyl tensor. It is worth pointing
out that a direct calculation of the transformed Weyl tensor for the
metric (\ref{transmet}) would be quite difficult.
The identities of section 2 will allow for an indirect 
approach to the result.

Let us start by evaluating the covariant derivative of
$\ssig_{\alpha}$ with respect to the transformed metric. To do that
it is convenient
to use the following identity, which is a trivial consequence
of the vanishing of the torsion of a Levi-Civita connection,
\begin{eqnarray*}
\nablapr_{\alpha} \ssig_{\beta} = \frac{1}{2} \left (
\nabla_{\alpha} \ssig_{\beta} - \nabla_{\beta} \ssig_{\alpha} \right )
+ \frac{1}{2} \pounds_{\vec{\ssig}} \g _{\alpha\beta},
\end{eqnarray*}
where $\vec{\ssig}$ is the vector obtained by raising the indices of
$\ssig_{\alpha}$ with the metric $\g_{\alpha\beta}$, i.e.
\begin{eqnarray}
\ssig^{\alpha} \equiv \left [\g^{-1} \right ]^{\alpha\beta} \ssig_{\beta} = 
\frac{1}{\left ( \delta + i \gamma \sigma \right )^2 \nn^2}
\left [\sigma^{\beta} + \xi^{\beta} \left (W^{\mu} \sigma_{\mu} \right )
\right ].
\label{transigmavec}
\end{eqnarray}
Using that $\vec{\xi}$ is a Killing vector for the metric
$\g_{\alpha\beta}$, a not-too-long calculation gives 
\begin{eqnarray}
\nablapr_{\alpha} \ssig_{\beta} = 
\nabla_{\alpha} \left [ \frac{\sigma_{\beta}}{\left (\delta + i \gamma \sigma
\right)^2} \right ] - \frac{1}{\nn^2 \left (
\delta + i \gamma \sigma \right )^2} \left \{ \frac{}{}
\sigma_{\left ( \alpha \right .}
\nabla_{\left . \beta \right )} \nn^2 
+ \sigma^{\mu} \left [ \frac{}{} 2 F_{\mu \left ( \alpha \right .}
W_{\left . \beta \right )} \right . \right . \nonumber  \\ 
\left . \left . \frac{}{} - 4 \gamma \mbox{Re} \left [
\left (\gamma \ov{\sigma} + i \delta \right ) \F_{\mu \left ( \alpha
\right . } \right ] \left (\xi_{\left . \beta \right )} 
+ \frac{\lambda}{\nn^2} W_{\left . \beta \right )} \right )
+ \frac{1}{2} W_{\alpha} W_{\beta} \nabla_{\mu} \left ( \frac{\lambda}{\nn^2}
\right ) - \frac{1}{2} g_{\alpha\beta} \nabla_{\mu} \nn^2 
\right ] \right \},
\label{firstlong}
\end{eqnarray}
where, as usual, round brackets enclosing indices denote
symmetrization. Let us keep this
expression for later use and let us now evaluate $\nablapr_{\alpha} \xi^{\mu}
\Fpr_{\mu\beta}$. We start by
raising one index to $\Fpr_{\alpha\beta}$
with respect to the metric $\g_{\alpha\beta}$, i.e. we evaluate
$\Fpr^{\mu}_{\,\,\,\beta} \equiv 
[\g^{-1}]^{\mu\nu} \Fpr_{\nu\beta}$. The result reads
\begin{eqnarray*}
\Fpr^{\mu}_{\,\,\,\beta} =
\frac{1}{\left ( i \gamma \sigma + \delta \right )^2}
\left [ \F^{\mu}_{\,\,\,\beta} + \xi^{\mu} W_{\alpha} \F^{\alpha}_{\,\,\,
\beta} + \frac{1}{2 \nn^2} W_{\beta} \sigma^{\mu} 
+ \frac{1}{2 \nn^2} \xi^{\mu} W_{\beta} 
\sigma^{\rho} W_{\rho} \right ].
\end{eqnarray*}
We now take into account that $\nablapr_{\alpha} \xi^{\mu} \Fpr_{\mu\beta}
= \mbox{Re} \left ( \Fpr_{\alpha\mu} \right ) \Fpr^{\mu}_{\,\,\,\nu}$ so that
only the latter has to be calculated. A simple calculation gives
\begin{eqnarray}
\mbox{Re} \left ( \Fpr_{\alpha\mu} \right ) \Fpr^{\mu}_{\,\,\,\beta}
= \frac{1}{\left ( \delta + i \gamma \sigma \right )^2}
\mbox{Re} \left [ \frac{1}{\left ( \delta + i \gamma\sigma \right )^2}
\left (\F_{\alpha\mu}  - \frac{1}{2 \nn^2} W_{\alpha} \sigma_{\mu}
\right ) \right ]
\left [ \nn^2 \F^{\mu}_{\,\,\,\beta} + \frac{1}{2} W_{\beta} \sigma^{\mu}
\right ].
\label{intermedio1}
\end{eqnarray}
To proceed further we need to use two identities. 
The first one is valid for any complex quantity $B$ and reads
\begin{eqnarray}
\mbox{Re} \left ( \frac{B}{\left ( \delta + i \gamma \sigma \right )^2}
\right ) = \frac{1}{\nn^4} \left [
\left (\delta + i \gamma \sigma \right )^2 \mbox{Re}(B) +
2 B \gamma \lambda
\left ( \gamma \frac{\sigma - \ov{\sigma}}{2} - i \delta \right ) \right ],
\label{iden1}
\end{eqnarray}
which can be proven easily from the trivial relation
$\mbox{Re} (A B)  = \ov{A} \, \mbox{Re} (B)
+ B (A - \ov{A})/2$. The second identity reads
\begin{eqnarray}
\left (\F_{\alpha\mu} - \frac{1}{2 \nn^2 } W_{\alpha} \sigma_{\mu}
\right )
\left (\F^{\mu}_{\,\,\,\beta} + \frac{1}{2 \nn^2 } W_{\beta} \sigma^{\mu}
\right  ) = - \frac{\Fsq}{4 \nn^2} \g_{\alpha\beta},
\label{iden2}
\end{eqnarray}
and it is proven by expanding the left-hand side and using standard 
properties of self-dual 2-forms (see Sect. 2). 
Inserting (\ref{iden1}) and (\ref{iden2}) into (\ref{intermedio1}) we 
get
\begin{eqnarray}
\nablapr_{\alpha} \xi^{\mu} \Fpr_{\mu\beta} =
\frac{1}{\nn^2}
\mbox{Re} \left [ \left (\F_{\alpha\mu}  - \frac{1}{2 \nn^2} W_{\alpha}
\sigma_{\mu} \right ) \right ] 
\left ( \F^{\mu}_{\,\,\beta} + \frac{1}{2\nn^2} W_{\beta} \sigma^{\mu}
\right )- \hspace{3cm} \nonumber \\
\hspace{9cm}  - \frac{ \Fsq \gamma \lambda \left (\gamma \frac{\sigma - 
\ov{\sigma}}{2} - i \delta \right )}{2 \nn^4 \left (
\delta + i \gamma \sigma \right )^2} \g_{\alpha\beta}.
\label{secondlong}
\end{eqnarray}
We can now combine (\ref{firstlong}) and (\ref{secondlong})
in order to obtain  an explicit expression for
$\nablapr_{\alpha} \ssig_{\beta} - 2 \nablapr_{\alpha}
\xi^{\mu} \Fpr_{\mu\beta}$. This is achieved by using
$\nabla_{\mu} \lambda = \mbox{Re} (\sigma_{\mu})$, together
with the identity $\mbox{Re} [  (\gamma \ov{\sigma} + i \delta )
\F_{\alpha\beta} ] = (\gamma \sigma - i \delta) \mbox{Re} (\F_{\alpha\beta})
+ \F_{\alpha\beta} (i \delta + \gamma (\ov{\sigma} - \sigma )/2)$ (which is
proven in a similar way as (\ref{iden2})). Two further pieces of
information are required to obtain the result: identity
(\ref{qq1}) and the relation  $\sigma^{\mu} F_{\beta\mu} +
\nabla^{\mu} \lambda \F_{\mu\beta} = 0$, which
can be deduced from the last equality in (\ref{WW2}). 

The calculation now is rather involved. However, many cancellations happen
along the way and the final result turns out to be surprisingly simple.
It reads
\begin{eqnarray}
\nablapr_{\alpha} \ssig_{\beta} - 2 \nablapr_{\alpha}
\xi^{\mu} \Fpr_{\mu\beta} = \frac{1}{\left (\delta +
i \gamma \sigma \right )^2} \left ( \nabla_{\alpha} \sigma_{\beta} - 2
\nabla_{\alpha} \xi^{\mu} \F_{\mu\beta} \right ) - \hspace{4cm} \nonumber \\
\hspace{7cm}
- \frac{3 i \gamma}{\left ( \delta + i \gamma \sigma \right )^3}
\left [
\sigma_{\alpha} \sigma_{\beta} + \frac{1}{3} \Fsq \left (\lambda 
g_{\alpha\beta} + \xi_{\alpha} \xi_{\beta} \right ) \right ].
\label{final}
\end{eqnarray}
This expression is valid for any spacetime for which the spacetime
Ehlers group can be defined. No restriction to vacuum spacetimes
is necessary. Let us now restrict ourselves to the vacuum case and
let us define 
the symmetric and trace-free tensor
$\Y_{\alpha\beta} \equiv 2 \xi^{\mu} \xi^{\nu}
\C_{\mu\alpha\nu\beta}$. Since, in vacuum, $\R_{\alpha\beta\gamma\delta} =
\C_{\alpha\beta\gamma\delta}$ the transformation law of
$\Y_{\alpha\beta}$ can be obtained directly from
(\ref{derivsigma}) and (\ref{final}) to be
\begin{eqnarray}
\Ypr_{\alpha\beta} = \frac{\Y_{\alpha\beta}}{\left (\delta +
i \gamma \sigma \right )^2} 
- \frac{3 i \gamma}{\left ( \delta + i \gamma \sigma \right )^3}
\left [
\sigma_{\alpha} \sigma_{\beta} + \frac{1}{3} \Fsq \left (\lambda 
g_{\alpha\beta} + \xi_{\alpha} \xi_{\beta} \right ) \right ].
\label{tranW}
\end{eqnarray}
This formula is the key expression for obtaining
the transformation law for the full Weyl tensor. To do that, we first 
rewrite (\ref{tranW}) as follows,
\begin{eqnarray}
\xi^{\mu} \xi^{\nu} \C^{\prime}_{\mu\alpha\nu\beta} =
\frac{1}{\left ( \delta + i \gamma \sigma \right )^2}
\xi^{\mu} \xi^{\nu} \left [ \C_{\mu\alpha\nu\beta} 
- \frac{6 i \gamma}{\delta + i \gamma \sigma}
\left ( \F_{\mu\alpha} \F_{\nu\beta}- \frac{\Fsq}{3}
\I_{\mu\alpha\nu\beta} \right ) \right ],
\label{rewr}
\end{eqnarray}
where $\I_{\mu\alpha\nu\beta} \equiv
(g_{\mu\nu}\, g_{\alpha\beta}-g_{\mu\beta}\,g_{\alpha\nu}
 + i \,\eta_{\mu\alpha\nu\beta})/4$ is the canonical
metric in the
space of self-dual 2-forms. The term
is parenthesis in (\ref{rewr}) is self-dual in the metric $g_{\alpha\beta}$,
with respect to each pair of antisymmetric indices. 
On the other hand, $\C^{\prime}_{\mu\alpha\nu\beta}$ is self-dual
with respect to the metric $\g_{\alpha\beta}$. So, if we knew a method
to transform self-dual two-forms in $g_{\alpha\beta}$ into self-dual
two-forms in $\g_{\alpha\beta}$, we could obtain the full transformation
law for the Weyl tensor. This is addressed in the following Lemma
which is proven by straightforward, if somewhat long, calculation
\begin{lemma} Let $P^{\mu\nu}_{\alpha\beta}$ be defined as
$P^{\mu\nu}_{\alpha\beta} = \Omega^2 \delta^{\mu}_{\alpha}
\delta^{\nu}_{\beta} - \delta^{\mu}_{\alpha} \xi^{\nu} W_{\beta}
- \xi^{\mu} W_{\alpha} \delta^{\nu}_{\beta}$.
Then, a two-form ${\cal X}_{\alpha\beta}$ is self-dual
in $(\M,g)$ if and only if ${\cal X}^{\prime}_{\alpha\beta} \equiv
P^{\mu\nu}_{\alpha\beta} {\cal X}_{\mu\nu}$ is self-dual in
$(\M, \g)$.
\label{dualdual}
\end{lemma}
{\bf Remark.} This Lemma is true for any spacetime $(\M,g)$ and
for any pair $(\vec{\xi}, \bm{W} ) \in G$, not only for the 
spacetime Ehlers group we are considering in this section.

\vspace{2mm}

\noindent Thus, let us define the tensor
\begin{eqnarray*}
{\cal B}_{\alpha\beta\gamma\delta}
= \frac{1}{\left ( \delta + i \gamma \sigma \right)^2}
P^{\mu\nu}_{\alpha\beta} P^{\rho\sigma}_{\gamma\delta}
\left [ \C_{\mu\nu\rho\sigma}
 - \frac{6 i \gamma}{\delta + i \gamma \sigma}
\left ( \F_{\alpha\beta} \F_{\rho\sigma}- \frac{\Fsq}{3}
\I_{\alpha\beta\rho\sigma} \right ) \right ],
\end{eqnarray*}
which is, by construction, self-dual (for each pair of antisymmetric
indices) with respect to the metric $\g_{\alpha\beta}$. Furthermore, from 
$\xi^{\alpha} P^{\mu\nu}_{\alpha\beta} = \xi^{\mu} \left (
\delta^{\nu}_{\beta} - W_{\beta} \xi^{\nu} \right )$, we find
$\xi^{\mu} \xi^{\nu} \C^{\prime}_{\mu\alpha\nu\beta} =
\xi^{\mu} \xi^{\nu} {\cal B}_{\mu\alpha\nu\beta}$.
It is well-know, and an easy consequence of (\ref{xiF}), that,
for an arbitrary self-dual 2-form $\bm{\cal X}$ 
the following equation holds
\begin{eqnarray*}
2 \lambda {\cal X}_{\alpha\beta} =  \left (
\xi_{\beta} X_{\alpha} - \xi_{\alpha} X_{\beta} - i 
\eta_{\alpha\beta\rho\sigma} \xi^{\rho} X^{\sigma} \right ),
\end{eqnarray*}
where $X_{\beta} = 2 \xi^{\alpha} {\cal X}_{\alpha\beta}$.
A similar expression exists for objects with several pairs of antisymmetric
indices. Thus, we can conclude $\lambda ( \C^{\prime}_{\mu\alpha\nu\beta}
- {\cal B}_{\mu\alpha\nu\beta} ) = 0$. Using now that 
${\cal B}_{\mu\alpha\nu\beta}$ is continuous everywhere (including
the points where $\lambda =0$), we can conclude that, for
spacetimes where the Killing vector is null at most 
on a set with empty interior, $\C^{\prime}_{\mu\alpha\nu\beta}
= {\cal B}_{\mu\alpha\nu\beta}$. 
Summarizing, we have proven the following
\begin{proposition}
\label{propo}
Let $(\M,g)$ be a spacetime satisfying the hypotheses of Theorem \ref{vac}.
Assume further that the set of points where the Killing vector
is null has empty interior.
Then, the Weyl tensor of the spacetime $(\tilde{M},\T(g))$ reads
\begin{eqnarray}
\C^{\prime}_{\alpha\beta\gamma\delta}
= \frac{1}{\left ( \delta + i \gamma \sigma \right)^2}
P^{\mu\nu}_{\alpha\beta} P^{\rho\sigma}_{\gamma\delta}
\left [ \C_{\mu\nu\rho\sigma}
 - \frac{6 i \gamma}{\delta + i \gamma \sigma}
\left ( \F_{\mu\nu} \F_{\rho\sigma}- \frac{\Fsq}{3}
\I_{\mu\nu\rho\sigma} \right ) \right ],
\label{GenWeyl}
\end{eqnarray}
where $P^{\mu\nu}_{\alpha\beta}$ is defined in Lemma \ref{dualdual}
above.
\label{Propo}
\end{proposition}
{\bf Remark}.
The condition that $\vec{\xi}$ is non-null
almost everywhere can be shown to be unnecessary, i.e. that
Proposition \ref{Propo} {\it also} holds
for Killing vector with an arbitrary causal character. The sketch of the proof
is as follows. First obtain the transformation law
for the Weyl tensor under the linearized Ehlers group for an arbitrary
Killing vector. The solution turns out to be
the linearized version of (\ref{GenWeyl}). Then, the result follows
by exploiting the group structure of the spacetime Ehlers group.

The transformation law (\ref{GenWeyl}) for the Weyl tensor
is very simple indeed. In addition to
the necessary factors which transform self-dual objects in $g$ into
self-dual objects in $\g$, the essential part of the transformation is,
besides a global conformal factor $(\delta + i \gamma \sigma)^{-2}$,
adding to the original self-dual Weyl tensor a term proportional to
\begin{eqnarray*}
\left ( \F_{\mu\nu} \F_{\rho\sigma}- \frac{\Fsq}{3}
\I_{\mu\nu\rho\sigma} \right).
\end{eqnarray*}
This tensor is the simplest self-dual, symmetric and trace-free object that
can be constructed out of the Killing form. Taking into account that the
calculations leading to this proposition are quite long, the result is
surprisingly simple and elegant.
Since the curvature singularities of $(\M,\g)$ must be
singularities in $\C^{\prime}_{\mu\alpha\nu\beta}$,
expression (\ref{GenWeyl}) 
shows that, in general, curvature singularities in 
$(\M,\g)$ appear at the points where
$\delta+ i \gamma \sigma =0$ (i.e. where $\lambda = 0$ and $w= -\delta
/\gamma$). These points were precisely those that had to be excluded
from $\M$ in order to  define the spacetime Ehlers transformation.
No further singularities may appear. Thus, we
have a good control on the behaviour of the transformed spacetime without
having the perform the Ehlers transformation explicitly, which may be a
difficult task.

\section{A local characterization of the Kerr-NUT metric.}

The transformation law (\ref{GenWeyl}) for the Weyl tensor allows us to
select privileged subsets of solutions of the Einstein vacuum
field equations, namely those that remain invariant under the Ehlers 
group. The transformation law given in Proposition \ref{Propo}
suggests one of these invariant subsets.
Let us consider those vacuum spacetimes  admitting a Killing vector field
such that
\begin{eqnarray}
\C_{\alpha\beta\gamma\delta} =
Q \left (\F_{\alpha\beta} \F_{\gamma\delta} - \frac{1}{3}
\I_{\alpha\beta\gamma\delta} \Fsq \right )
\label{weylSi}
\end{eqnarray}
holds for some complex function $Q$.  Under an Ehlers transformation, the
tensor ${\cal N}_{\alpha\beta\gamma\delta}$  
retains the same form (with a different $Q$). If we take now into account
that the terms in parenthesis in (\ref{GenWeyl}) are just those necessary to 
transform $\bm{F}$ into $\bm{F'}$, the invariance  of the set of
solutions satisfying (\ref{weylSi}) follows easily. Actually, this is
the simplest possible invariant subset of the vacuum Einstein field equations.
It is remarkable that 
condition (\ref{weylSi}) appears also in a completely different
context, namely in a local characterization of the Kerr metric
obtained in \cite{Mars1}. In that paper the following result was proven
\begin{theorem}
\label{Teor1}
Let $(\V,g)$ be a smooth, vacuum spacetime admitting a Killing
vector $\vec{\xi}$. Assume that $(\V,g)$ is not locally flat and
that the following two conditions hold:

1. There exists at least one point $p \in \V$ such that
$\Fsq |_p \neq 0$.

2. The self-dual Weyl tensor and the Killing form associated to 
$\vec{\xi}$ satisfy (\ref{weylSi}).

Then, the Ernst one-form $\sigma_{\mu}$ is exact ( $\sigma_{\mu} =
 \nabla_{\mu} \sigma$) and $Q$ and $\Fsq$ must 
take the form $Q = -6 /(c - \sigma)$ , $\Fsq = A (c- \sigma)^4$ where
$A \neq 0 $ and $c$ are complex constants. 

If, in addition, $\mbox{Re} (c) > 0 $ and $A$ is real, then
the spacetime $(\V,g)$ is locally isometric to a Kerr spacetime.
\end{theorem}

This theorem was not explicitly stated in this form in \cite{Mars1}.
However, it is not difficult to see that the proof of the main theorem in
that paper also proves this theorem  (see \cite{Mars2} for a discussion).

This theorem suggests a natural question, namely 
which spacetimes correspond
to the other values of $A$ and $c$?. In this section we will answer
this question and will obtain a local characterization of the Kerr-NUT metric
by combining the theorem above and the action of the Ehlers group
discussed in this paper. The spacetimes satisfying the hypotheses
of \ref{Teor1} can be classified by the complex constants
$A \neq 0$ and $c$. Since, in addition, the whole family is invariant
under the Ehlers group (the fact that 
the Ehlers transformed $\Fprsq$ remains non-zero somewhere
follows from the transformation law (\ref{Fsqlaw})) we can
consider the action of the Ehlers group on the parameter
space defined by $A$ and $c$. Give that both 
the hypotheses and the conclusions of Theorem {\ref{Teor1} are
local, all the considerations below will also be local and there are no obstructions to define and apply Ehlers transformations.

Using the transformation law for the Weyl tensor (\ref{GenWeyl})
and the transformation law for $\Fsq$ (\ref{Fsqlaw}) we easily find
\begin{eqnarray}
\cpr = \frac{\alpha c + i \beta}{\delta + i \gamma c},
\hspace{2cm} 
\Apr = A \left (\delta + i \gamma c \right )^4,
\label{transAc}
\end{eqnarray}
where $\cpr$ and $\Apr$ are the corresponding values for the Ehlers
transformed spacetime. The parameter space defined by $A$ and $c$ is four-dimensional
and the Ehlers group has three real parameters. Thus, there must be one real
function of $A$ and $c$ which is invariant under Ehlers transformations.
It is easy to check that $A \ov{A} \left ( c + \ov{c} \right )^4$ fulfills
these requirements. In order to classify
the spacetimes satisfying the hypotheses of Theorem \ref{Teor1},
we need to find a unique representative of each orbit of the Ehlers
group in the space $(A, c)$. In order to do that, we notice that $c+ \ov{c}$ 
transforms as
\begin{eqnarray*}
\cpr + \ov{\cpr} = \frac{c + \ov{c}}{\left (\delta + i \gamma c \right )
\left ( \delta - i \gamma \ov{c} \right )}.
\end{eqnarray*}
Thus, $c + \ov{c}$ cannot change
sign under an Ehlers transformation (and it must remain zero if it was
originally zero). An easy inspection of the transformation law
(\ref{transAc}) shows that the orbits corresponding to $c +\ov{c}=0$
are uniquely characterized by the unit complex number $\frac{A}{\left | A
\right |}$ (where the vertical bars denote the norm of the complex number).
More precisely, any pair $c = i s_1$, $A = A_0 \exp{(i B)}$
(with
$s_1$, $A_0>0$ and $B$ being real) can be transformed into 
$\cpr= 0$ and  $\Apr = \exp{(iB)}$. 

Take now an arbitrary point $(A,c)$ with $\mbox{Re} (c) \neq 0$.
It is easy to check that there 
always exists an Ehlers transformation that brings this point
into the canonical form $\cpr = \mbox{sign} ( c + \ov{c}) $
and $\Apr = - \left | A \right  | \mbox{Re}(c)^2$ (we have chosen a negative
sign in $\Apr$ just for convenience, a positive sign can also be achieved).
So, the vacuum solutions satisfying the hypotheses
of Theorem \ref{Teor1} can be classified in three classes
according to the Ehlers group as follows
\begin{itemize}
\item Those with $\mbox{Re} (c) =0$, for which the orbit is determined
by $A/|A|$,
\item Those with $\mbox{Re} (c) >0$ and the orbit
is determined by the real constant $- \left | A \right | \mbox{Re} (c)^2$, 
\item  Those orbits with $\mbox{Re} (c) <0$ and the orbit is 
determined by $- \left | A \right | \mbox{Re} (c)^2$. 
\end{itemize}

With this classification at hand,
can now use Theorem \ref{Teor1} to obtain
a purely local geometric characterization of the Kerr-NUT spacetime. 
It is well-known that the Kerr-NUT family of spacetimes
is obtained and exhausted by applying the Ehlers transformations
to the Kerr spacetime.
Thus, the following theorem follows by combining the classification
discussed above and the results of Theorem \ref{Teor1}.

\begin{theorem}
\label{Teor2}
Let $(\V,g)$ satisfy the hypotheses of Theorem \ref{Teor1}.
If $\mbox{Re} (c) > 0$ then the spacetime $(\V,g)$ is locally
isometric to a Kerr-NUT spacetime.
\end{theorem}
This theorem extends a result by J.P. Krisch \cite{JPK} which finds the most general
vacuum solution on a strictly stationary spacetime which has vanishing Simon tensor
\cite{Simon}. The relationship between the vanishing of the Simon tensor and the
characterization of Kerr given in Theorem \ref{Teor1} is discussed in
detail in \cite{Mars1}.

We could still ask which spacetimes correspond to a zero or
a negative value of $\mbox{Re} (c)$. Without giving the proof, let us just
mention that they belong to the vacuum subset of the
Pleba\'nski limit of the rotating C metric
\cite{Cmetric}. They are analogous of the Kerr-NUT spacetime but
with the geometry of certain quotient (defined by the
stationary Killing vector and one of the principal null directions)
being not a round 2-sphere but a Euclidean plane
(when $\mbox{Re} (c) = 0$) or a Poincar\'e plane (when $\mbox{Re} (c) < 0$). 

\section*{Acknowledgements}

I would like to thank B.Schmidt, W.Simon and J.M.M. Senovilla for useful
comments on a previous version of this paper. 
This work has been partially supported by projects 
Oesterreichische Nationalbank Nr. 7942 and 
UPV172.310-G02/99.

\end{document}